\documentclass[a4paper,fleqn,usenatbib]{mnras}

\usepackage{newtxtext,newtxmath}

\usepackage[T1]{fontenc}
\usepackage{ae,aecompl}

\usepackage{graphicx}	
\usepackage{amsmath}	

\title[Origin of supermassive black holes in massive metal-poor protoclusters]{Origin of supermassive black holes in massive metal-poor protoclusters}

\author[Schleicher et al.]{
D.R.G. Schleicher$^{1}$\thanks{E-mail: dschleicher@astro-udec.cl},
B. Reinoso$^{2}$,
M. Latif$^{3}$,
R.S. Klessen$^{2, 4}$,
M.Z.C. Vergara$^{1}$,\newauthor
A. Das$^{5,6}$,
P. Alister$^{2}$,
V.B. D\'iaz$^{7}$, P.A. Solar$^1$
\\
$^{1}$ Departamento de Astronom\'ia, Facultad Ciencias F\'isicas y Matem\'aticas, Universidad de Concepci\'on, Av. Esteban Iturra s/n Barrio Universitario, Concepci\'on, Chile\\
$^{2}$ Universit\"at Heidelberg, Zentrum f\"ur Astronomie, Institut f\"ur theoretische Astrophysik, Albert-Ueberle Str. 2, 69120 Heidelberg, Germany\\
$^{3}$ Physics Department, College of Science, United Arab Emirates University, PO Box 15551, Al-Ain, UAE\\
$^4$ Universit\"at Heidelberg, Interdisziplin\"ares Zentrum f\"ur wissenschaftliches Rechnen, Im Neuenheimer Feld 205, 69120 Heidelberg, Germany\\
$^{5}$ Universities Space Research Association (USRA), Mountain View, CA 94043, USA\\
$^{6}$ NASA Ames Research Center (NASA ARC), Moffett Field, CA 94035, USA\\
$^7$ Hamburger Sternwarte, Universit\"at Hamburg, Gojenbergsweg 112, 21029 Hamburg, Germany
}

\date{Accepted XXX. Received YYY; in original form ZZZ}

\pubyear{2022}

\begin{document}
\label{firstpage}
\pagerange{\pageref{firstpage}--\pageref{lastpage}}
\maketitle

\begin{abstract}
While large numbers of supermassive black holes have been detected at $z>6$, their origin is still essentially unclear. Numerical simulations have shown that the conditions for the classical direct collapse scenario are very restrictive and fragmentation is very difficult to be avoided. We thus consider here a more general case of a dense massive protostar cluster at low metallicity ($\lesssim10^{-3}$~Z$_\odot$) embedded in gas. We estimate the mass of the central massive object, formed via collisions and gas accretion, considering the extreme cases of a logarithmically flat and a Salpeter-type initial mass function. Objects with masses of at least $10^4$~M$_\odot$ could be formed for inefficient radiative feedback, whereas $\sim10^3$~M$_\odot$ objects could be formed when the accretion time is limited via feedback. These masses will vary depending on the environment and could be considerably larger, particularly due to the continuous infall of gas into the cloud. As a result, one may form intermediate mass black holes of $\sim10^4$~M$_\odot$ or more. Upcoming observations with the James Webb Space Telescope (JWST) and other observatories may help to detect such massive black holes and their environment, thereby shedding additional light on such a formation channel.
\end{abstract}




\section{Introduction}
The possible formation pathways of supermassive black holes were originally proposed by \citet{Rees1984}. These included the formation via direct collapse \citep[e.g.][]{Bromm2003, Wise2008a, Koushiappas2004, Begelman06, Schleicher10b, Latif2013BH, LatifSchleicher2015, Regan2017, Grete2019, Suazo2019, Chon2020, Latif2021}, through collisions in dense stellar clusters \citep[e.g.][]{Poregies2002, Devecchi09, Katz2015, Sakurai2017, Reinoso2018, Escala2021, Vergara2021}, or more exotic scenarios involving mergers in clusters of stellar black holes \citep[e.g.][]{Davies2011, Lupi2014, Kroupa2020}. The formation of  massive black holes could occur directly or through the formation of other types of progenitor objects, such as supermassive stars  \citep[e.g.][]{Appenzeller1972a, Appenzeller1972b, Fuller1986, Begelman2010, Hosokawa2013, Schleicher2013, Umeda2016, Haemmerle2020, Haemmerle2021}.

While the original pathways outlined by \citet{Rees1984} were either based on stellar-dynamical or gas-dynamical mechanisms, recently it became more apparent that realistic scenarios will likely require a combination of the two. On the one hand, this is because fragmentation cannot be fully avoided even under ideal conditions when the gas is primordial and molecular hydrogen formation efficiently inhibited via soft-UV radiaton \citep[e.g.][]{Latif2013BH, Latif2020}, while on the other hand, it is very likely that molecular hydrogen is indeed present and will enhance fragmentation, as a very strong soft-UV background is required for its destruction \citep[][]{Latif2014a,Sugimura2014,Agarwal2015,Latif2015}. In the presence of heavy elements or dust grains, fragmentation effectively cannot be avoided \citep[e.g.][]{Schneider2006, Clark2008,Omukai2008, Schneider2010, Bovino2014, Peters2014, Grassi2017}, thus favoring the formation of a stellar cluster rather than a single object. 

On the other hand, the black hole masses formed via collisional scenarios were often found to be of the order $10^3$~M$_\odot$ \citep[e.g.][]{Devecchi2009, Reinoso2018}, while models indicate that the explanation of the observed supermassive black holes at high redshift \citep[e.g.][]{Inayoshi2020} may require more massive seeds \citep{Shapiro2005, Valiante2016, Sassano2021, Wang2021}. Also from other environments, i.e. the formation of very massive stars, it is well-known that gas-dynamical and stellar dynamical processes may interact and favor the formation of more massive objects \citep{Baumgardt2011}. 

Particularly, due to accretion, protostellar radii can be enhanced \citep{Hosokawa2012, Hosokawa2013, Schleicher2013}, thus increasing the probability for collisions. In the context of black hole growth, \citet{Alexander2014} considered a stellar mass black hole embedded in a metal-rich cluster consisting of gas and protostars, showing how a supra-exponential growth is possible during the early phase of its growth. The black hole formation scenario by \citet{Davies2011} and \citet{Lupi2014} similarly requires an inflow of gas into a cluster of stellar mass black holes, in order to sufficiently steepen the gravitational potential and make mergers more likely than three-body ejections. \citet{Boekholt2018} explored the interaction of collisions and accretion in a compact primordial cluster embedded in gas, showing that potentially masses of $\sim10^5$~M$_\odot$ can be reached. \citet{Alister2020} have extended this analysis including mass loss during mergers. \citet{Tagawa2020} developed a semi-analytic model to follow the growth of a supermassive object via stellar bombardment in the presence of gas, finding that the formation of a potential massive object depends on the densities where gas fragments. \citet{Chon2020} similarly have shown that dust cooling may not prevent efficient accretion, and thus is not necessarily prohibitive to the formation of massive objects. \citet{Das2021a} considered the impact of different analytic accretion rates and showed how these enhance the probability for collisions, whereas \citet{Das2021b} explored the evolution of nuclear star clusters embedded in gas, considering mass loss from stellar winds.

Overall, it is natural to explore such scenarios that involve gas and stellar clusters, due to the ubiquity of gas in the first galaxies and the expected formation of protoclusters. At the same time, due to the complexity of the problem and the multitude of the physics involved, it is hard to exhaustively treat the problem via numerical simulations, and we need to arrive at a more general understanding considering the processes at play, with the aim to derive approximate relations and expectations that can be compared to simulation results in the future. In this paper, we first consider the general properties and evolution of massive protoclusters in very metal-poor gas ($\lesssim10^{-3}$~Z$_\odot$) in the first galaxies, as found in numerical simulations \citep[e.g.][]{Wise2008a, Latif2013BH, Latif2021}. This involves an analysis of the protostellar and gas-dynamical timescales in section~\ref{cluster}, together with an assessment of the potential relevance of feedback. The formation of a massive central object via stellar and gas-dynamical processes is considered in section~\ref{blackhole}, along with its further evolution, leading to the formation of a massive black hole. We summarize and discuss this scenario in section~\ref{discussion}, along with implications for future observations for instance with the James Webb Space Telescope (JWST)\footnote{JWST: https://www.jwst.nasa.gov/}.

\section{Basic considerations for gas-rich protostellar clusters}\label{cluster}
We consider a cloud with a total mass of $\sim6\times10^4$~M$_\odot$ on a scale of $R=0.25$~pc, as found in simulations by \citet{Latif2016}. We here assume that a relevant fraction of $\sim33\%$ of the total mass has been converted into stars, thus implying a gas mass  $M_g=4\times10^4$~M$_\odot$ and a stellar mass $M_*=2\times10^4$~M$_\odot$. We consider this as a somewhat conservative scenario; if indeed a smaller fraction of the mass goes into stars, more of the gas will be directly available for accretion and the fragmentation problem might be less relevant. As in the simulations, we further assuming a continous ongoing inflow rate of  $\dot{M}_{\rm in}\sim1$~M$_\odot$~yr$^{-1}$ into the cloud. We further restrict us here to metallicities of $\lesssim10^{-3}$~Z$_\odot$, i.e. where metal-line cooling is not relevant \citep{Bromm2003}, in order fragmentation to occur on too large scales. For simplicity, we assume the overall distribution to be homogeneous and spherically symmetric.

\subsection{The initial mass function of the protostars}\label{IMF}

In the case of a primordial cluster, we expect a logarithmically flat initial mass function (IMF) \citep{Hartwig2015}, implying that\begin{equation}
	\frac{dN(M)}{d\log M}=C=\mathrm{const}\qquad\Rightarrow\qquad\frac{dN}{dM}=CM^{-1},
\end{equation}
where $N(M)$ is the number of stars with mass $M$. We assume that there is a minimum stellar mass $M_{\rm min}$ and a maximum stellar mass $M_{\rm max}$. As the total stellar mass $M_*$ is known, we have\begin{equation}
	M_*=\int_{M_{\rm min}}^{M_{\rm max}} \left( \frac{dN}{dM} \right)MdM=C\left(M_{\rm max}-M_{\rm min}\right),
\end{equation}
thus fixing the normalization constant $C$. The mean stellar mass is then given as\begin{equation}
	\langle M\rangle=\frac{\int \frac{dN}{dM}MdM}{\int \frac{dN}{dM}dM}=\frac{M_{\rm max}-M_{\rm min}}{\ln \left(M_{\rm max}\right)-\ln\left( M_{\rm min}\right)}.
\end{equation}
Assuming typical parameters such as $M_{\rm max}=100$~M$_\odot$ and $M_{\rm min}=0.1$~M$_\odot$ thus implies $\langle M\rangle\sim14.5$~M$_\odot$, or  $N\sim1.3\times10^3$ for the cluster considered here.  In case the IMF is not logarithmically flat, but follows a power-law of the form\begin{equation}
	\frac{dN}{dM}=C_2M^\alpha,
\end{equation}
where typically we expect $\alpha$ to be negative, the normalization constant $C_2$ is given from\begin{equation}
	M_*=\int_{M_{\rm min}}^{M_{\rm max}} \left( \frac{dN}{dM} \right)MdM=C_2\frac{M_{\rm max}^{\alpha+2}-M_{\rm min}^{\alpha+2}}{\alpha+2},
\end{equation}
and the mean stellar mass is given as\begin{equation}
	\langle M\rangle = \left(\frac{M_{\rm max}^{\alpha+2}-M_{\rm min}^{\alpha+2}}{M_{\rm max}^{\alpha+1}-M_{\rm min}^{\alpha+1}}\right)\left(\frac{\alpha+1}{\alpha+2}\right).
\end{equation}
Assuming a Salpeter IMF with $\alpha=-2.35$, it implies $\langle M\rangle\sim0.35$~M$_\odot$ and $N=5.7\times10^4$ protostars. We note that the presence of magnetic fields in these clusters is potentially relevant, and was shown to support the formation of a top-heavy IMF \citep{Sharda2021}, and also \citet{Latif22} found strong indications of a top-heavy IMF with radiation-hydrodynamical simulatios. In the following, we will consider both the logarithmically flat IMF and the Salpeter IMF as two possible extreme cases.

\subsection{Stellar crossing and relaxation time}\label{relax}
To analyze the dynamics of the protostellar cluster, we adopt the formulation of \citet{Reinoso2020}, who extended the framework of \citet{Spitzer1987} to include the effect of a gas potential. The crossing time of the cluster is then given as\begin{equation}
	t_{\rm cross}=\frac{R}{V}=4.4\times10^3\mathrm{\ years},
\end{equation}
with \begin{equation}
	V=\sqrt{\frac{GM_*}{R}}(1+q),
\end{equation}
and $q=M_g/M_*=2$ with the parameters adopted here, assuming virial equilibrium. The relaxation time is then given as\begin{equation}
	t_{\rm relax}=0.138\frac{N(1+q)^4}{\ln(\gamma N)}t_{\rm cross}=\Bigg \{ \begin{matrix} 1.1\times10^7\mathrm{\rm\ years} & {\rm log-flat}\\ 2.8\times10^8{\rm\ years} & {\rm Salpeter}  \end{matrix}
\end{equation}
where $\gamma\sim0.4$ is a parameter related to the ratio of maximum to minimum impact parameter within the system. We note that the relaxation time is considerably enhanced by the gravitational potential from the gas with a factor of $(1+q)^4=81$. We note that even decreasing $M_*$ would not lead to a reduction of the relaxation time, and while $N$ would decrease with $M_*$, this decrease would be overcompensated via the factor $(1+q)^4$. So no significant contraction of the cluster would be expected on a timescale shorter than $10$ million years. 

\subsection{The gas and dynamical friction}\label{friction}
We next consider the behavior of the gas within the cluster. We first calculate the free-fall time of the gas, given as \begin{equation}
	t_{\rm ff}=\sqrt{\frac{3\pi}{32G\rho}}\sim1.0\times10^4\mathrm{\ years},
	\end{equation}
where $\rho$ is the density of the gas, which we evaluate assuming spherical symmetry and a homogeneous distribution inside the cloud, i.e. \begin{equation}\rho=M_g/(4\pi R^3/3)\sim4.1\times10^{-17}\mathrm{\ g~cm}^{-3}.\end{equation}
Thus in principle, the gas could evolve and contract under a relatively short timescale. Under approximately virialized conditions, the kinetic and thermal energy of the gas must correspond to about half of  the gravitational energy, i.e.\begin{equation}
	\frac{GM_g(M_g+M_*)}{R}\sim M_g v_{\rm eff}^2,
\end{equation} 
where $v_{\rm eff}$ corresponds to the turbulent or thermal velocity, depending on which one is the larger component. Under the conditions considered here, we thus find $v_{\rm eff}\sim32$~km~s$^{-1}$. On the other hand, even under conditions where H$_2$ cooling is suppressed, the temperature at the densities considered here is expected to be $T\sim6000$~K, implying a sound speed of $c_s=\sqrt{\gamma k_B T/m}=8.3$~km~s$^{-1}$ for an ideal gas consisting of hydrogen and helium, with $k_B$ the Boltzmann constant, $\gamma=5/3$ the adiabatic index and $m=1.2m_p$, assuming a hydrogen-helium gas, with $m_p=1.67\times10^{-24}$~g the proton mass. This overall implies that the gas exhibits strongly supersonic turbulent motions, which should decay on a timescale \citep{MacLow1999}\begin{equation}
	t_{\rm decay}=\frac{R}{v_{\rm eff}}=7.6\times10^3\mathrm{\ years}.
\end{equation}
This timescale is comparable to the free-fall time of the gas, and as a result, free-fall could start after one turbulent decay time, unless the turbulent energy is sustained via some mechanism.

A way to replenish the turbulent energy in the gas is through the dyamical friction exerted by the stars onto the gas. Based on the formalism of \citet{Ostriker1999}, we have\begin{equation}
	\frac{dv}{dt}=-\frac{4\pi G^2M\rho}{c_s^2}f^{\rm (gas)},
\end{equation} 
with\begin{equation}
	f^{\rm (gas)}=\frac{1}{\mathcal{M}^2}\left[ \frac{1}{2}\ln(\mathcal{M}^2-1)+\ln\Lambda \right]
\end{equation}
in the case where the relative motion between the stars and the gas, expressed via the Mach number $\mathcal{M}=V/c_s$, is highly supersonic, i.e. $\mathcal{M}\gg1$. In this expression, $\ln\Lambda$ is the so-called Coulomb logarithm for the gas distribution, where we adopt $\ln\Lambda=3.1$ as suggested by \citet{Chapon2013}. Considering the stellar velocities estimated above, as well as the fact that the gas temperature cannot be higher than $6000$~K, we have $\mathcal{M}\geq6.7$, and $f^{\rm (gas)}\sim0.11$. The friction force acting on one star thus corresponds to $F\sim0.11\times4\pi G^2\langle M\rangle^2\rho/c_s^2$. Within one crossing time $t_{\rm cross}$, an average star will cross a spatial scale $R$ and the energy transferred from the stars into the gas thus corresponds to $FR$. Given the $N$ stars in the clusters, the total energy transfer in one crossing time corresponds to $NFR$, and the energy transfer rate is thus $E_{\rm dyn}\sim NFR/t_{\rm cross}=NFV$. A significant change of the kinetic energy of the gas could thus be expected on a timescale\begin{equation}
	t_{\rm df,g}=\frac{\frac{1}{2}M_gv_{\rm eff}^2}{NFV}=\Bigg \{ \begin{matrix} 5.6\times10^6\mathrm{\rm\ years} & {\rm log-flat}\\ 2.3\times10^8{\rm\ years} & {\rm Salpeter}  \end{matrix}
\end{equation}

While dynamical friction is not expected to have a significant effect on the energy budget of the gas, it may still have some effect on the stars themselves. Defining the characteristic timescales for stars to change their velocity by a significant degree as\begin{equation}
	t_{\rm df,*}=\frac{V}{\dot{v}}=\frac{V}{F/\langle M\rangle }=\Bigg \{ \begin{matrix} 1.7\times10^7\mathrm{\rm\ years} & {\rm log-flat}\\ 6.9\times10^8{\rm\ years} & {\rm Salpeter}  \end{matrix}\label{dynfric}
\end{equation}
These timescales are comparable to the relaxation time of the cluster and thus suggest that contraction of the cluster could be driven by the combination of both processes simultaneously. We also consider energy input by accretion-driven turbulence, as suggested by \citet{Klessen2010}. Numerical simulations, e.g. by \citet{Latif2013BH}, have shown accretion rates $\dot{M}\sim1$~M$_\odot$~yr$^{-1}$ and infall velocities $v_{\rm in}\sim10$~km~s$^{-1}$. The corresponding timescale to significantly change the kinetic energy of the cloud through the deposited kinetic energy is thus
\begin{equation}
	t_{\rm acc}=\frac{\frac{1}{2}M_gv_{\rm eff}^2}{\frac{1}{2}\dot{M}v_{\rm in}^2}\sim4.1\times10^5\mathrm{\ years}.
\end{equation}
As a result, the timescales for the generation of new turbulent energy thus appear considerably larger than the decay time by at least an order of magnitude. This implies that turbulence may only delay but not prohibit the collapse, thereby favoring the formation of a massive object.

\subsection{Protostellar accretion}\label{protoacc}
We now aim to understand the typical accretion of the protostars in the cluster, considering the presence of supersonic turbulence and surrounding protostars. Under conditions of spherical symmetry, the accretion of an object with mass $\langle M\rangle$ moving through the gas with a velocity $V$ is given by the Bondi solution \citep{Bondi1952}, which we write in the form given by \citet{Maccarone2012}:\begin{equation}
	\dot{M}_{\rm BH}=7\times10^{-9}\left(\frac{\langle M\rangle }{M_\odot} \right)^2\left( \frac{n}{10^6\mathrm{\ cm}^{-3}} \right)^2\left( \frac{\sqrt{c_s^2+V^2}}{10^6\mathrm{\ cm\ s}^{-1}} \right)^{-3}\ M_\odot\mathrm{\ yr}^{-1},\label{Bondi}
\end{equation}
with $n=\rho/\mu\sim2.1\times10^7$~cm$^{-3}$ being the number density of the gas, $\mu=1.2m_p$ the mean molecular weight for a neutral hydrogen-helium gas, and $c_s$ the sound speed evaluated at infinity (far away from the protostar). Since the medium considered here is supersonically turbulent, we will replace the sound speed with the effective velocity $v_{\rm eff}$ derived above. In addition, it was noted by \citet{Kaaz2019} that the separation between the stars should be taken into consideration when evaluation the accretion rates, particularly when the separation is smaller than the Bondi radius given as \begin{equation}
	R_B=\frac{G\langle M\rangle }{v_{\rm eff}^2}=\Bigg \{ \begin{matrix} 6.0\times10^{-5}\mathrm{\rm\ pc} & {\rm log-flat}\\ 1.5\times10^{-6}{\rm\ pc} & {\rm Salpeter}  \end{matrix}\qquad,
\end{equation}
where we considered the impact of supersonic turbulence on the accretion process. We can estimate the separation between the stars $R_\perp$ as \begin{equation}
	R_\perp \sim RN^{-1/3}=\Bigg \{ \begin{matrix} 0.022\mathrm{\rm\ pc} & {\rm log-flat}\\ 0.0065{\rm\ pc} & {\rm Salpeter}  \end{matrix}\qquad.
\end{equation} 
The separation thus appears large enough to not require corrections to the accretion rate. Considering the effective turbulent velocity instead of the sound speed, we thus expect a typical accretion rate of
\begin{equation}
	\dot{M}_{\rm BH}\sim\Bigg\{\begin{matrix} 2.3\times10^{-6}\mathrm{\ M}_\odot\mathrm{\ yr}^{-1} & \mathrm{log-flat}\\ 1.4\times10^{-9}\mathrm{\ M}_\odot\mathrm{\ yr}^{-1} & \mathrm{Salpeter}   \end{matrix}\qquad.
\end{equation}
 Evaluating the timescale to significantly change the mass of a typical star, we find
 
 \begin{equation}
 	 \langle M\rangle/\dot{M}_{\rm BH}=\Bigg\{\begin{matrix}  6.2\times10^6\mathrm{\ years} & \mathrm{log-flat} \\2.5\times10^8\mathrm{\ years} & \mathrm{Salpeter}  \end{matrix}\qquad.
 \end{equation}
Individual stars may nonetheless grow faster due to the steep dependence of the accretion rate on the protostellar mass rate as $M^2$. We also check on the timescale for the gas to be depleted due to accretion, \begin{equation} t_{\rm dep}=\frac{M_g}{N\dot{M}_{\rm BH}}=\Bigg \{ \begin{matrix} 1.2\times10^7\mathrm{\rm\ years} & {\rm log-flat}\\ 5.1\times10^8{\rm\ years} & {\rm Salpeter}  \end{matrix}\qquad.
\end{equation} 
With inflow rates of $\dot{M}_{\rm in}\sim1$~M$_\odot$~yr$^{-1}$, the typical inflow time is $M_g/\dot{M}_{\rm in}=4\times10^4$~years, implying a significant mass increase during the evolution of the cluster.

\begin{table}
	\centering
	\begin{tabular}{ccc}\hline
		timescale [years]    &  log-flat & Salpeter \\ \hline
		crossing time & $4.4\times10^3$ & $4.4\times10^3$\\
		turbulence decay time & $7.6\times10^3$ & $7.6\times10^3$\\
		free-fall time & $1.0\times10^4$ & $1.0\times10^4$\\
		mass infall timescale & $4\times10^4$ & $4\times10^4$\\
		gas evaporation timescale & $4\times10^4$ & $1\times10^6$  \\
		turbulent energy accretion timescale & $4.1\times10^5$ & $4.1\times10^5$\\
		Bondi timescale protostar & $6.2\times10^6$ & $2.5\times10^8$\\
		depletion timescale through protostellar accretion & $1.2\times10^7$ & $5.1\times10^8$\\
		relaxation time & $1.1\times10^7$ & $2.8\times10^8$\\
		dynamical friction time of stars due to gas & $1.7\times10^7$ & $6.9\times10^8$\\
	\hline
	\end{tabular}
	\caption{A comparison of the characteristic timescales of the cluster for the logarithmically flat and the Salpeter IMF{, ordered from the shortest to the longest timescales.}}
	\label{timescales}
\end{table}

\subsection{Implications of characteristic timescales: gravity vs. feedback}
In section~\ref{friction}, we found that the timescale for dynamical friction is comparable to the timescale for dynamical relaxation. As derived above, the characteristic timescale of typical protostars to change their masses is then even shorter. A significant change of the typical stellar masses can be expected to have further implications for the evolution of the cluster, as the accretion of mass will likely also imply the accretion of linear and angular momentum. Due to the supersonic turbulence in the cluster, the accreted linear and angular momentum will correspond (approximately) to a random walk. The accreted angular momentum is unlikely to be aligned with the previous angular momentum of the protostars with respect to the center, and at least some of the stars are then expected to sink towards the center due to the reduced angular momentum, increasing the overall likelihood for collisions. 

Independent of this, a steepening of the cluster is also expected due to the continuing mass inflow to the center. At least initially, the cluster is still exposed to an inflow rate of $\dot{M}_{\rm in}\sim1$~M$_\odot$~yr$^{-1}$ \citep{Latif2016}, and even over longer timescales up to $\sim7\times10^3$~years, cosmological radiation hydrodynamics simulations found inflow rates in the range of $10^{-2}-10^{-1}$~M$_\odot$~yr$^{-1}$ by \citep{Latif2021}. The increase of the total mass implies a steepening of the gravitational potential, implying a contraction of the cluster. We estimate the latter via adiabatic contraction \citep{Blumenthal1986}, implying that\begin{equation}
	M(r)r=\mathrm{constant},
\end{equation}
with $M(r)$ being the total mass enclosed in the mass scale $r$. While the simulations by \citet{Latif2021} found some variations of the gas infall rate over time, these did not show very systematic trends but rather stochastic variations, and we can thus at least very roughly assume an approximately constant inflow rate $\dot{M}_{\rm in}$ over the first million years or so. This implies that the mass will evolve as\begin{equation}
	M_{\rm tot}=M_{\rm ini}+\dot{M}_{\rm in}\Delta t,
\end{equation}
where $\Delta t$ is the time passed after the cluster had a mass of $M_{\rm ini}$ with a size of $R_{\rm ini}$. The size of the cluster is thus expected to evolve as\begin{equation}
	R(\Delta t)=R_{\rm ini}\frac{M_{\rm ini}}{M_{\rm ini}+\dot{M}_{\rm in}\Delta t}.
\end{equation}
Particularly once $\dot{M}_{\rm in}\Delta t\gg M_{\rm ini}$, we then have\begin{equation}
	R(\Delta t)\propto \left( \dot{M}_{\rm in}\Delta t \right)^{-1}.
\end{equation}
Each time the cluster mass doubles due to infall, the radius is thus expected to contract at least by a factor of $2$. This will likely be a lower limit, due to the presence of additional physical processes, such as the effect of the accretion onto the protostars (implying the more heavy ones to sink to the center) and the effect of dynamical friction.

Such a steepening only occurs if the continuous infall into the cluster is not counter-acted via another energy source. In the case of the Salpeter IMF, the mean stellar mass is so low and  the characteristic timescale for the stars to grow is so large that feedback is not expected to be relevant for the typical stars. Considering the luminosities provided by \citet{Windhorst2018}, the stellar luminosity over mass ratio becomes significant roughly starting from $10$~M$_\odot$ stars. For the Salpeter IMF, the number of stars more massive than at $10$~M$_\odot$ is given as\begin{equation}
	\int_{10\ M_\odot}^{M_{\rm max}}\left(\frac{dN}{dM}  \right)dM=\frac{C_2}{\alpha+1}\left(M_{\rm max}^{\alpha+1}-(10\ M_\odot)^{\alpha+1}  \right).
\end{equation}
The fraction of stars more massive than $10$~M$_\odot$ follows as\begin{equation}
	\epsilon_{\rm mass}=\frac{\int_{10\ M_\odot}^{M_{\rm max}}\left(\frac{dN}{dM}  \right)dM}{\int_{M_{\rm min}}^{M_{\rm max}}\left(\frac{dN}{dM}  \right)dM}=\frac{M_{\rm max}^{\alpha+1}-(10\ M_\odot)^{\alpha+1}}{M_{\rm max}^{\alpha+1}-M_{\rm min}^{\alpha+1}}\sim1.9\times10^{-3}
\end{equation}
with the parameters adopted in section~\ref{IMF}. We estimate the timescale to evaporate the cloud via the luminosity of massive stars as \begin{equation}
	t_{\rm evap}=\frac{E_{\rm grav}}{N\epsilon_{\rm mass} L_{10\ M_\odot}\epsilon_{\rm rad}},\label{evap}
\end{equation} 
where $E_{\rm grav}=GM_g(M_g+M_*)/R)\sim1.2\times10^{51}$~erg denotes the gravitational energy of the gas within the potential provided by the stars and the gas and $L_{10\mathrm{\ M}_\odot}=10^{3.86}L_\odot$ \citep{Windhorst2018}. The parameter $\epsilon_{\rm rad}$ describes the efficiency by which the energy produced in the protostellar source is deposited within the surrounding medium. For a warm ionized medium with density $0.1$~cm$^{-3}$, \citet{Haid2018} have shown that the efficiency rapidly drops from initially around $50\%$ to values between $10^{-4}$ and $10^{-5}$, while it is $\sim10^{-4}$ for a cold neutral medium at a density of $100$~cm$^{-3}$. The typical densities in our cluster are $\sim2\times10^7$~cm$^{-3}$, so to be conservative we assume an efficiency parameter of $10^{-2}$. With these assumptions, we find $t_{\rm evap}\sim10^6$~years, still considerably larger than the mass infall timescale $M_g/\dot{M}_{\rm in}\sim4\times10^4$~years.

It is less obvious in case of the logarithmically flat IMF, at least in case that the upper limit of the protostellar mass is considered to be $M_{\rm max}=100$~M$_\odot$, as adopted here. For the mean protostellar mass derived above, \citet{Windhorst2018} provide a luminosity of $L_*\sim2\times10^4$~L$_\odot$. We estimate the evaporation time via Eq.~(\ref{evap}) finding that $t_{\rm evap}\sim4\times10^4$~years. 

At the densities considered here, we further note that the HII regions around individual stars may get trapped and recombine due to the shorter recombination times at high density \citep[see also][]{Rahner2017, Rahner2019, Latif2021}. A rough estimate of their size can be obtained from the Str\"omgren radius \citep{Stromgren1939}\begin{equation}
	R_S=\left( \frac{3}{4\pi}\frac{S_*}{n^2\beta_2}, \right)^{1/3},
\end{equation} 
where $S_*$ is the number of ionizing photons per unit time and $\beta_2(T)=2\times10^{-10}T^{-3/4}$~cm$^3\,$s$^{-1}$ the recombination rate. For a star with $1000$~M$_\odot$, where we expect $S_*=1.6\times10^{50}$~s$^{-1}$ based on \citet{Bromm2001} and assuming $T=10^4$~K, we obtain $R_S=2.6\times10^{-3}$~pc for the densities ($n\sim2\times10^7$~cm$^{-3}$) on the $0.25$~pc scale. This is considerably below the typical separation of the protostars for the logarithmically flat IMF and also still below the mean separation for the Salpeter IMF. We further recall that the typical HII region of the protostars will be smaller, as here we calculated the extreme case with the upper mass limit of the IMF.

In future studies, it will be important to better quantify the efficiency parameter for the regime considered here. We note that the corresponding problem  arises for the logarithmically flat IMF if protostars are relatively massive at an early stage, while the problem would not be as relevant if the maximum mass of the IMF was $10$~M$_\odot$ or less, or in case of a Salpeter-type IMF, where the mean stellar mass is considerably reduced. An overview of the different timescales derived in this section is given in table~\ref{timescales}. In the following, we will consider the further evolution, considering the possibilities of inefficient and efficient radiative feedback. After longer times, i.e. a few million years, additional complications such as supernova feedback may arise. In the following, we will however restrict our discussion to the earlier stages.

\section{Formation and evolution of a central massive object}\label{blackhole}
In this section, we discuss whether a very massive protostar could be formed in a cluster as considered above, and eventually collapse into a massive black hole. Particularly, we aim to assess its formation via collisional and gas-related processes. {\citet{Regan2020} performed cosmological simulations for primordial halos exposed to a moderate Lyman Werner background, finding in one of their halos the formation of a very massive object with several thousand solar masses, and in a second halo a very massive object of a few hundred solar masses. While resolutions of the order $10^{-3}$~pc in principle are very good for cosmological simulations, they still limit any final conclusions that could be drawn. 
	
	We here} consider at least initially similar arguments as \citet{Devecchi09} and \citet{Katz2019} for a pure stellar cluster, which will subsequently be adapted for the presence of gas. We note here that several of our estimates likely are lower limits, as the protostars may be accreting, implying the presence of dynamical friction as well as the absence of angular momentum conservation, which overall should favor to enhance the probability of stars migrating to the center and participating in merger events. 

For a system of equal mass stars, the timescale for core collapse corresponds to \citep{Cohn1980} \begin{equation}
	t_{\rm cc}\sim15t_{\rm relax}\sim\Bigg\{\begin{matrix} 10^8\mathrm{\ years} & \mathrm{log-flat}\\ \gtrsim10^9\mathrm{years} & \mathrm{Salpeter}  \end{matrix}
\end{equation}
We consider these timescales here to be too large to be relevant. However, our cluster does not consist of equal mass stars, but follows an IMF. The stars are exposed to dynamical friction, and the massive stars will sink to the center. The core-collapse time in the cluster is then dictated by the time it takes for the most massive star to sink to the center, and can be estimated as \citep{Poregies2002}\begin{equation}
	t_{\rm df}=3.3\frac{\langle M\rangle}{M_{\rm max}}t_{\rm relax}=\Bigg\{\begin{matrix} 5.2\times10^6\mathrm{\ years} & \mathrm{log-flat}\\ 3.2\times10^6\mathrm{years} & \mathrm{Salpeter}  \end{matrix}
\end{equation}
in a pure stellar-dynamical scenario. Evaluating Eq.~(\ref{dynfric}) for the maximum stellar mass rather than the average stellar mass for the IMF, we can estimate the dynamical friction timescale due to the gas, finding $t_{\rm max,df}\sim1.4\times10^6$~years. Now, this star does not only experience dynamical friction but also accretes, and we can estimate the accretion timescale of such a star as $t_{\rm acc,max}=M_{\rm max}/\dot{M}_{\rm BH,max}=8.9\times10^5$~years. On this timescale, we further expect it to change its angular and linear momentum due to accretion. We in principle expect both processes to contribute, and thus estimate its infall time via \begin{equation}t_{\rm in}^{-1}=t_{\rm acc,max}^{-1}+t_{\rm max,df}^{-1}\sim5.4\times10^5\mathrm{\ years}.\end{equation}
To estimate the mass growth of the central object via collisions, we adopt the analytical model of \citet{Poregies2002} given as\begin{equation}
M_{\rm CMO}=M_{\rm ini}+4\times10^{-3}f_cM_*\ln\Lambda_c\left[\ln\left(\frac{t_{\rm diss}}{t_{c}}\right)+\frac{t_c}{t_{\rm diss}}-1  \right],\label{masscoll}
\end{equation}
where $M_{\rm ini}$ is the initial mass of the massive object, we recall that $M_*$ is the total stellar mass of the cluster, $f_c$ the fraction of binaries going through mergers, $\ln(\Lambda_C)$ the Coulomb logarithm as defined via \citet{Binney1987}, $t_{\rm diss}$ the dissolution timescale for the cluster, and $t_{c}$ the timescale on which core-collapse occurs in the cluster. We adopt the parametrization for the Coulomb logarithm from \citet{Poregies2002} as \begin{equation}
	\ln \left(\Lambda_C\right)=\ln\left( \frac{0.1M_*}{\langle M\rangle} \right).
\end{equation}
It is encouraging to note that this runaway-process appears not to significantly depend on the presence of rotation or the geometry of the cluster \citep{Vergara2021}. \citet{Poregies2002} obtained Eq.~\ref{masscoll} by analytically integrating over a term proportional to $\frac{1}{t}-\frac{1}{t_{\rm diss}}$, with $t_c$ and $t_{\rm diss}$ the integration limits (assuming collisions to start once the most massive star has arrived at the center due to dynamical friction and subsequently to last until $t_{\rm diss}$). We will generalize this here, as due to stellar evolution effects, the massive stars within the cluster will explode as supernovae after a timescale $t_{\rm SN}$. Changing the time limits of integration to be  $t_c$ and $t_{\rm SN}$, we obtain the new expression
\begin{equation}
	M_{\rm CMO,SN}=M_{\rm ini}+4\times10^{-3}f_cM_*\ln\Lambda_c\left[\ln\left(\frac{t_{\rm SN}}{t_{c}}\right)+\frac{t_c}{t_{\rm diss}}-\frac{t_{\rm SN}}{t_{\rm diss}}  \right],\label{masscoll2}
\end{equation}
\citet{Schaerer2002} derived an estimated lifetime of $2.5-3$~Myr for a $100$~M$_\odot$ Pop. III star. For a conservative estimate, we adopt here a lower limit of $t_{SN}\sim1$~Myr for supernova feedback to become relevant. We estimate $t_{\rm diss}\sim t_{\rm relaxx|}$, and thus $t_{\rm diss}\gg t_{\rm SN}$. In the pure $N$-body simulations of \citet{Poregies2002}, only a fraction $f_c\sim0.2$ of all binaries undergoes mergers. In a gaseous environment, we expect this fraction to be considerably enhanced due to the gas dynamical friction as well as the accretion torque due to the accretion of the gas onto the protostars. While it is hard to predict exactly what the right fraction should be, we adopt here a somewhat more optimistic value of $f_c\sim0.8$. The core collapse time $t_c$ is estimated here as $t_c=t_{\rm in}$. Evaluating Eq.~\ref{masscoll2} with this input, we obtain\begin{equation}
	M_{\rm CMO,SN}=\Bigg \{ \begin{matrix} 1.1\times10^3\mathrm{\rm\ M}_\odot & {\rm log-flat}\\ 1.8\times10^3{\rm\ M}_\odot & {\rm Salpeter}  \end{matrix} \qquad,\label{massSN}
\end{equation}
Now, it is important to take into account additional effects of the gas. First, we expect that the central massive object itself continues  to accrete gas while going through collisions. And second, also the protostars merging with the central object keep accreting until they merge, thereby contributing to an overall increase of the mass. Inspired by Eq.~(\ref{massSN}) we take a fixed mass of $1000$~M$_\odot$ for the evaluation of further mass growth by Bondi accretion. Using Eq.~(\ref{Bondi}), we find $\dot{M}_{\rm BH}\sim1.1\times10^{-2}$~M$_\odot$~yr$^{-1}$ or an accretion timescale $M_{\rm max}/\dot{M}_{\rm BH}\sim8.9\times10^3$~years. In principle this implies a significant growth via accretion, which could potentially grow as $\dot{M}_{\rm BH}\propto M^2$. In practice, however, this will be limited; on the one hand due to the effect noted by \citet{Maccarone2012} as nearby protostars would perturb the Bondi acccretion once the Bondi radius becomes too close. In the case of a logarithmically flat IMF, this would occur at roughly $M=1.5\times10^4$~M$_\odot$, and at $1.5\times10^3$~M$_\odot$ in case of the Salpeter IMF.

On the other hand, also the presence of any initial angular momentum in the gas could become a limiting factor to a certain extend. We note here, however, that in principle numerical simulations e.g. by \citet{Wise2008a} or \citet{Latif2013BH} did not find the structures on these scales to be strongly influenced by rotation. \citet{Alexander2014} already discussed in the context of a stellar mass black hole embedded in a turbulent cloud how the random walk of the black hole due to relaxation with the protostars in the cloud provides a path to effectively overcome the momentum barrier. This will be easier when the central massive object is still an extended protostellar object, increasing the overall cross section with the gas. The ambient medium here fulfills the condition that was outlined by \citet{Inayoshi2016} to permit hyper-Eddington accretion, i.e. we have\begin{equation}
	\left( \frac{n}{10^5\mathrm{\ cm}^{-3}} \right)\geq \left( \frac{M}{10^4\ M_\odot} \right)^{-1}\left( \frac{T}{10^4\ K} \right)^{3/2},
\end{equation}
with $T$ the temperature of the cloud (which will self-regulate at $\sim10^4$~K in case of photo-ionizing radiation, or otherwise be lower due to the radiative cooling of the gas). Based on these overall considerations, we generalize Eq.~(\ref{masscoll2}) to\begin{eqnarray}
	M_{\rm CMO,gas}&=&M_{\rm ini}+\dot{M}_{BH}t_{\rm gas}+4\times10^{-3}f_c M_*\ln\Lambda_c\nonumber\\
	&\times&\left[\ln\left(\frac{t_{\rm SN}}{t_{c}}\right)+\frac{t_c}{t_{\rm diss}}-\frac{t_{\rm SN}}{t_{\rm diss}}  \right],\label{masscollacc}
\end{eqnarray}
where we introduced a gas accretion term over the timescale $t_{\rm gas}$ when gas is available. For $t_{\rm gas}$, we consider an optimistic scenario where feedback is inefficient and thus $t_{\rm gas}\sim 10^6$~years, as well as a a scenario where feedback is more efficient, with $t_{\rm gas}\sim10^4$~years.  With $t_{\rm gas}=10^6$~years, we find
\begin{equation}
	M_{\rm CMO,gas,10^6}=\Bigg \{ \begin{matrix} 1.1\times10^4\mathrm{\rm\ M}_\odot & {\rm log-flat}\\ 1.2\times10^4{\rm\ M}_\odot & {\rm Salpeter}  \end{matrix}\qquad .
\end{equation}
These masses should still be lower limits, as we did not consider here the continous infall of mass into the cluster, which should lead to contraction and enhance the collapse and accretion. With $t_{\rm gas}=10^4$~years¸we obtain a mass of
\begin{equation}
	M_{\rm CMO,gas,10^4}=\Bigg \{ \begin{matrix} 1.2\times10^3\mathrm{\rm\ M}_\odot & {\rm log-flat}\\ 1.9\times10^3{\rm\ M}_\odot & {\rm Salpeter}  \end{matrix}\qquad.
\end{equation}

In the first case where radiative feedback is not relevant, we thus derived masses of the central object of at least $\sim10^4$~M$_\odot$, while they remain close to $\sim10^3$~M$_\odot$ in the presence of feedback. {In a super-competitive accretion scenario as outlined by \citet{Chon2020}, there could be a change of the balance between direct gas accretion by the CMO versus indirect effects, such as gas accretion by the protostars that subsequently merge with the most massive object. One can potentially expect that such direct gas accretion, when feasible, would be even more beneficial, given that not all of the protostars might merge with the central massive object during the available time. It will be important in future work to further establish the conditions when and for long how such super-competitive accretion is feasible. Even if it is not, our considerations here show that supermassive objects could be nonetheless formed.}

\begin{figure}
	\includegraphics[scale=0.35]{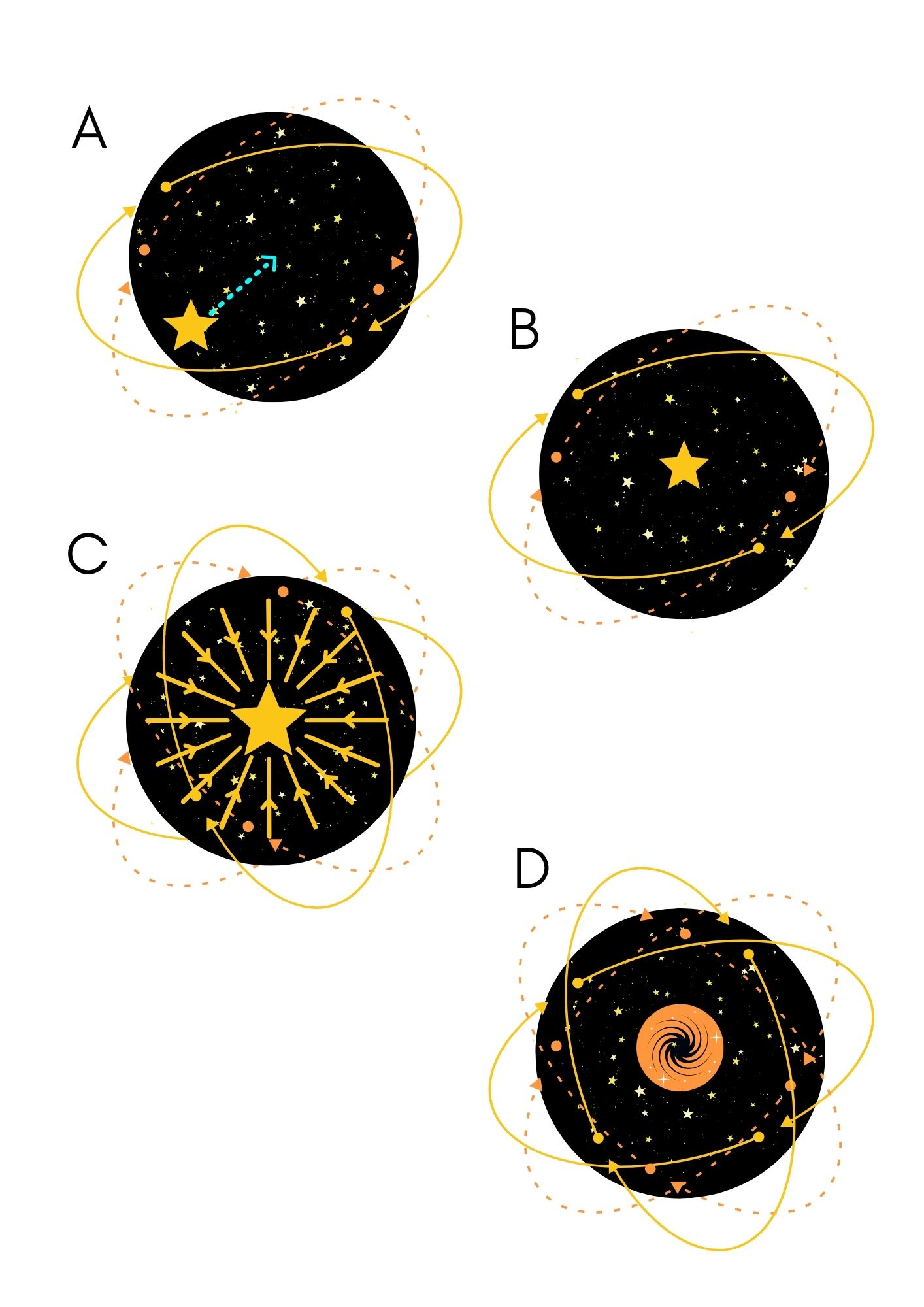}
	\caption{A sketch of the expected evolution of the stellar cluster and its central massive object. A-B) The most massive star falls to the center as a result of dynamical friction. C) The most massive star in the center accretes material through stellar collisions and from the gas. D) The supermassive star collapses into a very massive black hole.\label{scenario} }
\end{figure}

We noted above that Bondi accretion implies a typical accretion rate of $\sim10^{-2}$~M$_\odot$~yr$^{-1}$, and from the growth due to collisions, a similar average mass growth rate can be expected. Based on protostellar evolution calculations \citep[e.g.][]{Hosokawa2013, Schleicher2013, Umeda2016}, these rates  are too low to maintain a very extended envelope of the protostar, and it is more likely to contract on the Kelvin-Helmholtz timescale and form a supermassive star \citep[see also][]{Janka2002,Begelman2010}. Such objects are expected to subsequently collapse due to the general relativistic instability, as described by \citet{Appenzeller1972a, Fuller1986, Haemmerle2020, Haemmerle2021}. A summary sketch of the overall scenario considered here is given in Fig.~\ref{scenario}.

We  note that the numbers derived here are rough estimates, which may vary depending on the environment, such as the inflow rates for example, or the mass and compactness of the cluster. This is hard to quantify exactly, but one should certainly expect an environmental dependence of at least a factor of a few \citep[see also][]{Tagawa2020}. In reality, one thus needs to consider a range of possible outcomes in terms of mass. The upper limits for the masses of supermassive stars are considered to be in the range $10^{5}-10^{6}$~M$_\odot$ \citep{Woods2017, Haemmerle2018, Haemmerle2019}, providing an indication about the maximum seed masses that could be formed. In a small mass window, \citet{Chen2014} found that Pop.~III stars with $\sim55,000$~M$_\odot$ can cause thermonuclear explosions releasing energies of up to $10^{55}$~erg. This question regarding was recently reinvestigated by \citet{Nagele2020}, who did not find an explosion in the non-rotating case, though they found it when they included slow rotation. Such occasional bursts may thus occur, but are likely not the generic outcome. 

The case of rotating supermassive stars was investigated by \citet{Uchida2017}, finding that an additional torus can be formed surrounding the rotating black hole that forms as the outcome of collapse. Such studies were recently extended to incorporate the effect of magnetic fields, showing how collapse may lead to the launching of jets  consistent with the typical durations of long gamma-ray bursts \citep{Butler2018, Sun2018}. Observationally, this could be a possibly interesting implication, as it links this formation scenario to already known phenomena. Once a massive black hole is formed, it may continue to grow via tidal disruption events (TDEs) disrupting the surrounding stars \citep{Sakurai2019}. Such events may give rise to additional possibilities to detect the newly formed black holes.

\section{Discussion and conclusions}\label{discussion}

We have considered here the evolution of a compact and massive cloud consisting of gas and stars at very low metallicity. Our results strongly suggest that the interaction between the gas and the stars can lead to the formation of very massive central objects. In case that radiative feedback is not too relevant, as it could be the case for the Salpeter IMF, we find masses of the central object of at least $10^4$~M$_\odot$, which are lower limits as the cluster is exposed to continuous infall and contraction, which should further enhance the accretion and collapse. If the accretion timescale is limited by feedback, we found final masses of about $10^3$~M$_\odot$, which are then largely due to stellar collisions with gas accretion playing only a minor role. The relevance and efficiency of feedback needs to be further explored in these environments; estimates of the Str\"omgren radius show that the radiation could be trapped and might thus not be too prohibitive for the growth of the massive object. Beyond that, the masses in general are likely to depend on the environment, with higher mass in the presence of larger infall rates, or more massive and compact clusters \citep[see also][]{Tagawa2020}. Due to the low metallicity, the feedback from winds is not expected to be relevant, as sugested, e.g. by  \citet{Das2021b}. We conclude that the overall conditions within the cloud are expected to allow very efficient growth of the central massive object, both due to collisions in the dense cluster, but also through accretion in the gas. We particularly note that the presence of supersonic turbulence as well as the random walk of the central massive object within the cluster will highly alleviate the role of angular momentum, as also noted by \citet{Alexander2014}. The conditions here in the cloud further allow for the possibility of hyper-Eddington accretion, matching the condition derived by \citet{Inayoshi2016}.

Considering the masses estimated here and the expected dependence on the environment, we expect the formation of massive black holes with at least $10^{3}-10^{4}$~M$_\odot$, though possibly even reaching masses of $10^{5}-10^{6}$~M$_\odot$, depending on the environment and considering the upper limits derived by \citet[e.g.][]{Woods2017}. Occasionally if such stars have masses close to $55,000$~M$_\odot$, highly energetic explosions may also occur \citep{Chen2014, Nagele2020}, though we do not expect this as a very generic outcome. Producing massive black holes seeds was found to be important to explain the observed supermassive black holes at high redshift \citep{Shapiro2005, Valiante2016, Sassano2021}, and the formation mechanism outlined here may thus help to alleviate this problem.

To evaluate the impact of such initial masses, it will be particularly valuable to observationally study the low end of the black hole mass function at cosmic dawn \citep{Trinca2022}. The proposed formation scenario leads to several natural predictions that can be verified with future observations. A very natural prediction is that early black holes should be surrounded by a young star cluster, potentially even still embedded in gas, in case that feedback was inefficient. At the early stages and in very massive clouds, even strong obscuration could occur in these environments, potentially requiring sub-mm investigations with ALMA\footnote{ALMA: https://www.almaobservatory.org/en/home/} or hard X-ray studies with the future Athena X-ray Observatory\footnote{Athena: https://www.the-athena-x-ray-observatory.eu/} to verify the presence of the supermassive black hole. These will affect the mass function of supermassive black holes, as calculated by \citet{Sassano2021}, where the low-mass range is particularly relevant. The formation of these black holes from supermassive stars may be accompanied by the formation of energetic jets, with properties similar to known gamma-ray bursts \citep{Butler2018, Sun2018}. As the black hole will be initially embedded into a star cluster, the occurence and growth via TDEs is further a very much expected ingredient \citep{Sakurai2019}. {A possible and relevant concern has been raised by \citet{Pfister2019}, who explored the dynamics of black hole seeds in high-redshift galaxies, finding that it is dominated by the stellar component in case of seed masses less than $10^5$~M$_\odot$, which prevents
	 the black hole to sink to the center. It is currently unclear if this concern also applies to the formation mechanism that is considered here; particularly we note that the formation of a dense stellar cluster embedded in gas as envisioned here already requires the presence of a steep gravitational potential, which may help stabilize the very massive object in the place where it formed. A relevant factor may also concern the question on how quickly it subsequently grows, either via tidal disruption events or gas accretion. The question regarding the subsequent evolution will thus be of high relevance.}

It is  worth noting that the scenario proposed here requires different preconditions than e.g. black hole formation via direct collapse, where fragmentation needs to be strongly suppressed and it is typically assumed that molecular hydrogen should be fully destroyed and no metals or dust grains should be present. Thus, the formation sites are likely to be different in both scenarios, which is important to take into account to assess the further implications e.g. for galaxy evolution or to predict the expected black hole mass functions \citep{Habouzit2021, Trinca2022}. The scenario proposed here may certainly be influenced by metallicity, which changes the cooling and fragmentation behavior, affecting gas dynamics, fragmentation and the resulting IMF of the protostars. As shown here in this draft, this may also strongly influence the relevance of radiation feedback. Particularly if the metallicity is too high, it may become prohibitive if fragmentation occurs too early, and the cluster may thus not be compact enough. We expect this to be a problem particularly in the metal-line cooling regime \citep{Bromm2003}, while in the presence of dust cooling high accretion rates seem to be sustainable \citep{Chon2020, Chon2021}, thus requiring metallicities $\lesssim10^{-3}$~Z$_\odot$. A topic we did not discuss here in depth is the role of magnetic fields, though work e.g. by \citet{Hirano2021} and \citet{Sharda2021} has shown that magnetic fields can influence and increase the final masses in primordial environments. The efficient amplification of magnetic fields in halos where the first black holes could form has been demonstrated e.g. by \citet{Latif2013mag, Latif2014mag}, and indeed even in high Mach number flows magnetic field amplification can be relevant \citep{Schleicher2013mag, Schober2015}, and are likely to further alleviate any problems related to the angular momentum barrier. The presence of such fields will of course be highly favorable for jet-launching during the collapse of supermassive stars  \citep{Sun2018}. The inclusion of such magnetic effects will therefore likely give rise to additional phenomena and may thus favor the formation of very massive objects.

\section*{Acknowledgements}
DRGS gratefully acknowledges support by the ANID BASAL projects ACE210002 and FB210003 as well as via the Millenium Nucleus NCN19-058 (TITANs). DRGS thanks for funding via Fondecyt Regular (project code 1201280). BR acknowledges support through ANID (CONICYT-PFCHA/Doctorado acuerdo
bilateral DAAD/62180013) as well as support from DAAD (funding program
number 57451854). MAL thanks UAEU for funding via UPAR grant No. 31S390. PA acknowledges support through ANID (CONICYT-PFCHA/Doctorado acuerdo bilateral DAAD/62190003) as well as support from DAAD (DAAD/BECAS Chile funding program number 57504858). VBD acknowledges financial support from ANID (ANID-PFCHA/DOCTORADO DAAD-BECAS CHILE/62200025) as well as financial support from DAAD (DAAD/Becas Chile funding program ID 57559515). RSK acknowledges support from the Deutsche Forschungsgemeinschaft (DFG) via the Collaborative Research Center (SFB 881, Project-ID 138713538) 'The Milky Way System' (subprojects A1, B1, B2, B8) and from the Heidelberg cluster of excellence (EXC 2181 - 390900948) “STRUCTURES: A unifying approach to emergent phenomena in the physical world, mathematics, and complex data”. RSK also thanks for funding form the European Research Council in the ERC synergy grant “ECOGAL – Understanding our Galactic ecosystem: From the disk of the Milky Way to the formation sites of stars and planets” (project ID 855130).  MZCV and PAS thank for financial support from the Millenium Nucleus NCN19$\_$058 (TITANs).

\section*{Data availability}
All data relevant for this article are included in the article itself. Any potentially missing information can be requested from the first author via e-mail.




\bibliographystyle{mnras}
\bibliography{astro} 

\begin{thebibliography}{}
\makeatletter
\relax
\def\mn@urlcharsother{\let\do\@makeother \do\$\do\&\do\#\do\^\do\_\do\%\do\~}
\def\mn@doi{\begingroup\mn@urlcharsother \@ifnextchar [ {\mn@doi@}
  {\mn@doi@[]}}
\def\mn@doi@[#1]#2{\def\@tempa{#1}\ifx\@tempa\@empty \href
  {http://dx.doi.org/#2} {doi:#2}\else \href {http://dx.doi.org/#2} {#1}\fi
  \endgroup}
\def\mn@eprint#1#2{\mn@eprint@#1:#2::\@nil}
\def\mn@eprint@arXiv#1{\href {http://arxiv.org/abs/#1} {{\tt arXiv:#1}}}
\def\mn@eprint@dblp#1{\href {http://dblp.uni-trier.de/rec/bibtex/#1.xml}
  {dblp:#1}}
\def\mn@eprint@#1:#2:#3:#4\@nil{\def\@tempa {#1}\def\@tempb {#2}\def\@tempc
  {#3}\ifx \@tempc \@empty \let \@tempc \@tempb \let \@tempb \@tempa \fi \ifx
  \@tempb \@empty \def\@tempb {arXiv}\fi \@ifundefined
  {mn@eprint@\@tempb}{\@tempb:\@tempc}{\expandafter \expandafter \csname
  mn@eprint@\@tempb\endcsname \expandafter{\@tempc}}}

\bibitem[\protect\citeauthoryear{{Agarwal} \& {Khochfar}}{{Agarwal} \&
  {Khochfar}}{2015}]{Agarwal2015}
{Agarwal} B.,  {Khochfar} S.,  2015, \mn@doi [\mnras] {10.1093/mnras/stu1973},
  \href {https://ui.adsabs.harvard.edu/abs/2015MNRAS.446..160A} {446, 160}

\bibitem[\protect\citeauthoryear{{Alexander} \& {Natarajan}}{{Alexander} \&
  {Natarajan}}{2014}]{Alexander2014}
{Alexander} T.,  {Natarajan} P.,  2014, \mn@doi [Science]
  {10.1126/science.1251053}, \href
  {https://ui.adsabs.harvard.edu/abs/2014Sci...345.1330A} {345, 1330}

\bibitem[\protect\citeauthoryear{{Alister Seguel}, {Schleicher}, {Boekholt},
  {Fellhauer}  \& {Klessen}}{{Alister Seguel} et~al.}{2020}]{Alister2020}
{Alister Seguel} P.~J.,  {Schleicher} D.~R.~G.,  {Boekholt} T.~C.~N.,
  {Fellhauer} M.,   {Klessen} R.~S.,  2020, \mn@doi [\mnras]
  {10.1093/mnras/staa456}, \href
  {https://ui.adsabs.harvard.edu/abs/2020MNRAS.493.2352A} {493, 2352}

\bibitem[\protect\citeauthoryear{{Appenzeller} \& {Fricke}}{{Appenzeller} \&
  {Fricke}}{1972a}]{Appenzeller1972a}
{Appenzeller} I.,  {Fricke} K.,  1972a, \aap, \href
  {https://ui.adsabs.harvard.edu/abs/1972A&A....18...10A} {18, 10}

\bibitem[\protect\citeauthoryear{{Appenzeller} \& {Fricke}}{{Appenzeller} \&
  {Fricke}}{1972b}]{Appenzeller1972b}
{Appenzeller} I.,  {Fricke} K.,  1972b, \aap, \href
  {https://ui.adsabs.harvard.edu/abs/1972A&A....21..285A} {21, 285}

\bibitem[\protect\citeauthoryear{{Baumgardt} \& {Klessen}}{{Baumgardt} \&
  {Klessen}}{2011}]{Baumgardt2011}
{Baumgardt} H.,  {Klessen} R.~S.,  2011, \mn@doi [\mnras]
  {10.1111/j.1365-2966.2011.18258.x}, \href
  {https://ui.adsabs.harvard.edu/abs/2011MNRAS.413.1810B} {413, 1810}

\bibitem[\protect\citeauthoryear{{Begelman}}{{Begelman}}{2010}]{Begelman2010}
{Begelman} M.~C.,  2010, \mn@doi [\mnras] {10.1111/j.1365-2966.2009.15916.x},
  \href {https://ui.adsabs.harvard.edu/abs/2010MNRAS.402..673B} {402, 673}

\bibitem[\protect\citeauthoryear{{Begelman}, {Volonteri}  \& {Rees}}{{Begelman}
  et~al.}{2006}]{Begelman06}
{Begelman} M.~C.,  {Volonteri} M.,   {Rees} M.~J.,  2006, \mn@doi [MNRAS]
  {10.1111/j.1365-2966.2006.10467.x}, \href
  {http://adsabs.harvard.edu/abs/2006MNRAS.370..289B} {370, 289}

\bibitem[\protect\citeauthoryear{{Binney} \& {Tremaine}}{{Binney} \&
  {Tremaine}}{1987}]{Binney1987}
{Binney} J.,  {Tremaine} S.,  1987, {Galactic dynamics}

\bibitem[\protect\citeauthoryear{{Blumenthal}, {Faber}, {Flores}  \&
  {Primack}}{{Blumenthal} et~al.}{1986}]{Blumenthal1986}
{Blumenthal} G.~R.,  {Faber} S.~M.,  {Flores} R.,   {Primack} J.~R.,  1986,
  \mn@doi [\apj] {10.1086/163867}, \href
  {https://ui.adsabs.harvard.edu/abs/1986ApJ...301...27B} {301, 27}

\bibitem[\protect\citeauthoryear{{Boekholt}, {Schleicher}, {Fellhauer},
  {Klessen}, {Reinoso}, {Stutz}  \& {Haemmerl{\'e}}}{{Boekholt}
  et~al.}{2018}]{Boekholt2018}
{Boekholt} T.~C.~N.,  {Schleicher} D.~R.~G.,  {Fellhauer} M.,  {Klessen} R.~S.,
   {Reinoso} B.,  {Stutz} A.~M.,   {Haemmerl{\'e}} L.,  2018, \mn@doi [\mnras]
  {10.1093/mnras/sty208}, \href
  {https://ui.adsabs.harvard.edu/abs/2018MNRAS.476..366B} {476, 366}

\bibitem[\protect\citeauthoryear{{Bondi}}{{Bondi}}{1952}]{Bondi1952}
{Bondi} H.,  1952, \mn@doi [\mnras] {10.1093/mnras/112.2.195}, \href
  {https://ui.adsabs.harvard.edu/abs/1952MNRAS.112..195B} {112, 195}

\bibitem[\protect\citeauthoryear{{Bovino}, {Grassi}, {Schleicher}  \&
  {Latif}}{{Bovino} et~al.}{2014}]{Bovino2014}
{Bovino} S.,  {Grassi} T.,  {Schleicher} D.~R.~G.,   {Latif} M.~A.,  2014,
  \mn@doi [\apjl] {10.1088/2041-8205/790/2/L35}, \href
  {https://ui.adsabs.harvard.edu/abs/2014ApJ...790L..35B} {790, L35}

\bibitem[\protect\citeauthoryear{{Bromm} \& {Loeb}}{{Bromm} \&
  {Loeb}}{2003}]{Bromm2003}
{Bromm} V.,  {Loeb} A.,  2003, \mn@doi [\apj] {10.1086/377529}, \href
  {https://ui.adsabs.harvard.edu/abs/2003ApJ...596...34B} {596, 34}

\bibitem[\protect\citeauthoryear{{Bromm}, {Kudritzki}  \& {Loeb}}{{Bromm}
  et~al.}{2001}]{Bromm2001}
{Bromm} V.,  {Kudritzki} R.~P.,   {Loeb} A.,  2001, \mn@doi [\apj]
  {10.1086/320549}, \href
  {https://ui.adsabs.harvard.edu/abs/2001ApJ...552..464B} {552, 464}

\bibitem[\protect\citeauthoryear{{Butler}, {Lima}, {Baumgarte}  \&
  {Shapiro}}{{Butler} et~al.}{2018}]{Butler2018}
{Butler} S.~P.,  {Lima} A.~R.,  {Baumgarte} T.~W.,   {Shapiro} S.~L.,  2018,
  \mn@doi [\mnras] {10.1093/mnras/sty834}, \href
  {https://ui.adsabs.harvard.edu/abs/2018MNRAS.477.3694B} {477, 3694}

\bibitem[\protect\citeauthoryear{{Chapon}, {Mayer}  \& {Teyssier}}{{Chapon}
  et~al.}{2013}]{Chapon2013}
{Chapon} D.,  {Mayer} L.,   {Teyssier} R.,  2013, \mn@doi [\mnras]
  {10.1093/mnras/sts568}, \href
  {https://ui.adsabs.harvard.edu/abs/2013MNRAS.429.3114C} {429, 3114}

\bibitem[\protect\citeauthoryear{{Chen}, {Heger}, {Woosley}, {Almgren},
  {Whalen}  \& {Johnson}}{{Chen} et~al.}{2014}]{Chen2014}
{Chen} K.-J.,  {Heger} A.,  {Woosley} S.,  {Almgren} A.,  {Whalen} D.~J.,
  {Johnson} J.~L.,  2014, \mn@doi [\apj] {10.1088/0004-637X/790/2/162}, \href
  {https://ui.adsabs.harvard.edu/abs/2014ApJ...790..162C} {790, 162}

\bibitem[\protect\citeauthoryear{{Chon} \& {Omukai}}{{Chon} \&
  {Omukai}}{2020}]{Chon2020}
{Chon} S.,  {Omukai} K.,  2020, \mn@doi [\mnras] {10.1093/mnras/staa863}, \href
  {https://ui.adsabs.harvard.edu/abs/2020MNRAS.494.2851C} {494, 2851}

\bibitem[\protect\citeauthoryear{{Clark}, {Glover}  \& {Klessen}}{{Clark}
  et~al.}{2008}]{Clark2008}
{Clark} P.~C.,  {Glover} S. C.~O.,   {Klessen} R.~S.,  2008, \mn@doi [\apj]
  {10.1086/524187}, \href
  {https://ui.adsabs.harvard.edu/abs/2008ApJ...672..757C} {672, 757}

\bibitem[\protect\citeauthoryear{{Cohn}}{{Cohn}}{1980}]{Cohn1980}
{Cohn} H.,  1980, \mn@doi [\apj] {10.1086/158511}, \href
  {https://ui.adsabs.harvard.edu/abs/1980ApJ...242..765C} {242, 765}

\bibitem[\protect\citeauthoryear{{Das}, {Schleicher}, {Leigh}  \&
  {Boekholt}}{{Das} et~al.}{2021a}]{Das2021a}
{Das} A.,  {Schleicher} D. R.~G.,  {Leigh} N. W.~C.,   {Boekholt} T. C.~N.,
  2021a, \mn@doi [\mnras] {10.1093/mnras/stab402}, \href
  {https://ui.adsabs.harvard.edu/abs/2021MNRAS.503.1051D} {503, 1051}

\bibitem[\protect\citeauthoryear{{Das}, {Schleicher}, {Basu}  \&
  {Boekholt}}{{Das} et~al.}{2021b}]{Das2021b}
{Das} A.,  {Schleicher} D. R.~G.,  {Basu} S.,   {Boekholt} T. C.~N.,  2021b,
  \mn@doi [\mnras] {10.1093/mnras/stab1428}, \href
  {https://ui.adsabs.harvard.edu/abs/2021MNRAS.505.2186D} {505, 2186}

\bibitem[\protect\citeauthoryear{{Davies}, {Miller}  \& {Bellovary}}{{Davies}
  et~al.}{2011}]{Davies2011}
{Davies} M.~B.,  {Miller} M.~C.,   {Bellovary} J.~M.,  2011, \mn@doi [\apjl]
  {10.1088/2041-8205/740/2/L42}, \href
  {https://ui.adsabs.harvard.edu/abs/2011ApJ...740L..42D} {740, L42}

\bibitem[\protect\citeauthoryear{{Devecchi} \& {Volonteri}}{{Devecchi} \&
  {Volonteri}}{2009a}]{Devecchi09}
{Devecchi} B.,  {Volonteri} M.,  2009a, \mn@doi [\apj]
  {10.1088/0004-637X/694/1/302}, \href
  {http://adsabs.harvard.edu/abs/2009ApJ...694..302D} {694, 302}

\bibitem[\protect\citeauthoryear{{Devecchi} \& {Volonteri}}{{Devecchi} \&
  {Volonteri}}{2009b}]{Devecchi2009}
{Devecchi} B.,  {Volonteri} M.,  2009b, \mn@doi [\apj]
  {10.1088/0004-637X/694/1/302}, \href
  {https://ui.adsabs.harvard.edu/abs/2009ApJ...694..302D} {694, 302}

\bibitem[\protect\citeauthoryear{{Escala}}{{Escala}}{2021}]{Escala2021}
{Escala} A.,  2021, \mn@doi [\apj] {10.3847/1538-4357/abd93c}, \href
  {https://ui.adsabs.harvard.edu/abs/2021ApJ...908...57E} {908, 57}

\bibitem[\protect\citeauthoryear{{Fuller}, {Woosley}  \& {Weaver}}{{Fuller}
  et~al.}{1986}]{Fuller1986}
{Fuller} G.~M.,  {Woosley} S.~E.,   {Weaver} T.~A.,  1986, \mn@doi [\apj]
  {10.1086/164452}, \href
  {https://ui.adsabs.harvard.edu/abs/1986ApJ...307..675F} {307, 675}

\bibitem[\protect\citeauthoryear{{Grassi}, {Bovino}, {Haugb{\o}lle}  \&
  {Schleicher}}{{Grassi} et~al.}{2017}]{Grassi2017}
{Grassi} T.,  {Bovino} S.,  {Haugb{\o}lle} T.,   {Schleicher} D.~R.~G.,  2017,
  \mn@doi [\mnras] {10.1093/mnras/stw2871}, \href
  {https://ui.adsabs.harvard.edu/abs/2017MNRAS.466.1259G} {466, 1259}

\bibitem[\protect\citeauthoryear{{Grete}, {Latif}, {Schleicher}  \&
  {Schmidt}}{{Grete} et~al.}{2019}]{Grete2019}
{Grete} P.,  {Latif} M.~A.,  {Schleicher} D.~R.~G.,   {Schmidt} W.,  2019,
  \mn@doi [\mnras] {10.1093/mnras/stz1568}, \href
  {https://ui.adsabs.harvard.edu/abs/2019MNRAS.487.4525G} {487, 4525}

\bibitem[\protect\citeauthoryear{{Habouzit} et~al.,}{{Habouzit}
  et~al.}{2021}]{Habouzit2021}
{Habouzit} M.,  et~al., 2021, \mn@doi [\mnras] {10.1093/mnras/stab496}, \href
  {https://ui.adsabs.harvard.edu/abs/2021MNRAS.503.1940H} {503, 1940}

\bibitem[\protect\citeauthoryear{{Haemmerl{\'e}}}{{Haemmerl{\'e}}}{2020}]{Haemmerle2020}
{Haemmerl{\'e}} L.,  2020, \mn@doi [\aap] {10.1051/0004-6361/202039828}, \href
  {https://ui.adsabs.harvard.edu/abs/2020A&A...644A.154H} {644, A154}

\bibitem[\protect\citeauthoryear{{Haemmerl{\'e}}}{{Haemmerl{\'e}}}{2021}]{Haemmerle2021}
{Haemmerl{\'e}} L.,  2021, \mn@doi [\aap] {10.1051/0004-6361/202140893}, \href
  {https://ui.adsabs.harvard.edu/abs/2021A&A...650A.204H} {650, A204}

\bibitem[\protect\citeauthoryear{{Haemmerl{\'e}}, {Woods}, {Klessen}, {Heger}
  \& {Whalen}}{{Haemmerl{\'e}} et~al.}{2018}]{Haemmerle2018}
{Haemmerl{\'e}} L.,  {Woods} T.~E.,  {Klessen} R.~S.,  {Heger} A.,   {Whalen}
  D.~J.,  2018, \mn@doi [\mnras] {10.1093/mnras/stx2919}, \href
  {https://ui.adsabs.harvard.edu/abs/2018MNRAS.474.2757H} {474, 2757}

\bibitem[\protect\citeauthoryear{{Haemmerl{\'e}}, {Meynet}, {Mayer}, {Klessen},
  {Woods}  \& {Heger}}{{Haemmerl{\'e}} et~al.}{2019}]{Haemmerle2019}
{Haemmerl{\'e}} L.,  {Meynet} G.,  {Mayer} L.,  {Klessen} R.~S.,  {Woods}
  T.~E.,   {Heger} A.,  2019, \mn@doi [\aap] {10.1051/0004-6361/201936716},
  \href {https://ui.adsabs.harvard.edu/abs/2019A&A...632L...2H} {632, L2}

\bibitem[\protect\citeauthoryear{{Haid}, {Walch}, {Seifried}, {W{\"u}nsch},
  {Dinnbier}  \& {Naab}}{{Haid} et~al.}{2018}]{Haid2018}
{Haid} S.,  {Walch} S.,  {Seifried} D.,  {W{\"u}nsch} R.,  {Dinnbier} F.,
  {Naab} T.,  2018, \mn@doi [\mnras] {10.1093/mnras/sty1315}, \href
  {https://ui.adsabs.harvard.edu/abs/2018MNRAS.478.4799H} {478, 4799}

\bibitem[\protect\citeauthoryear{{Hartwig}, {Bromm}, {Klessen}  \&
  {Glover}}{{Hartwig} et~al.}{2015}]{Hartwig2015}
{Hartwig} T.,  {Bromm} V.,  {Klessen} R.~S.,   {Glover} S. C.~O.,  2015,
  \mn@doi [\mnras] {10.1093/mnras/stu2740}, \href
  {https://ui.adsabs.harvard.edu/abs/2015MNRAS.447.3892H} {447, 3892}

\bibitem[\protect\citeauthoryear{{Hirano}, {Machida}  \& {Basu}}{{Hirano}
  et~al.}{2021}]{Hirano2021}
{Hirano} S.,  {Machida} M.~N.,   {Basu} S.,  2021, \mn@doi [\apj]
  {10.3847/1538-4357/ac0913}, \href
  {https://ui.adsabs.harvard.edu/abs/2021ApJ...917...34H} {917, 34}

\bibitem[\protect\citeauthoryear{{Hosokawa}, {Omukai}  \& {Yorke}}{{Hosokawa}
  et~al.}{2012}]{Hosokawa2012}
{Hosokawa} T.,  {Omukai} K.,   {Yorke} H.~W.,  2012, \mn@doi [\apj]
  {10.1088/0004-637X/756/1/93}, \href
  {https://ui.adsabs.harvard.edu/abs/2012ApJ...756...93H} {756, 93}

\bibitem[\protect\citeauthoryear{{Hosokawa}, {Yorke}, {Inayoshi}, {Omukai}  \&
  {Yoshida}}{{Hosokawa} et~al.}{2013}]{Hosokawa2013}
{Hosokawa} T.,  {Yorke} H.~W.,  {Inayoshi} K.,  {Omukai} K.,   {Yoshida} N.,
  2013, \mn@doi [\apj] {10.1088/0004-637X/778/2/178}, \href
  {https://ui.adsabs.harvard.edu/abs/2013ApJ...778..178H} {778, 178}

\bibitem[\protect\citeauthoryear{{Inayoshi}, {Haiman}  \&
  {Ostriker}}{{Inayoshi} et~al.}{2016}]{Inayoshi2016}
{Inayoshi} K.,  {Haiman} Z.,   {Ostriker} J.~P.,  2016, \mn@doi [\mnras]
  {10.1093/mnras/stw836}, \href
  {https://ui.adsabs.harvard.edu/abs/2016MNRAS.459.3738I} {459, 3738}

\bibitem[\protect\citeauthoryear{{Inayoshi}, {Visbal}  \& {Haiman}}{{Inayoshi}
  et~al.}{2020}]{Inayoshi2020}
{Inayoshi} K.,  {Visbal} E.,   {Haiman} Z.,  2020, \mn@doi [\araa]
  {10.1146/annurev-astro-120419-014455}, \href
  {https://ui.adsabs.harvard.edu/abs/2020ARA&A..58...27I} {58, 27}

\bibitem[\protect\citeauthoryear{{Janka}}{{Janka}}{2002}]{Janka2002}
{Janka} H.-T.,  2002, in {Gilfanov} M.,  {Sunyeav} R.,   {Churazov} E.,  eds,
  Lighthouses of the Universe: The Most Luminous Celestial Objects and Their
  Use for Cosmology. p.~357 (\mn@eprint {arXiv} {astro-ph/0202028}),
  \mn@doi{10.1007/10856495\_56}

\bibitem[\protect\citeauthoryear{{Kaaz}, {Antoni}  \& {Ramirez-Ruiz}}{{Kaaz}
  et~al.}{2019}]{Kaaz2019}
{Kaaz} N.,  {Antoni} A.,   {Ramirez-Ruiz} E.,  2019, \mn@doi [\apj]
  {10.3847/1538-4357/ab158b}, \href
  {https://ui.adsabs.harvard.edu/abs/2019ApJ...876..142K} {876, 142}

\bibitem[\protect\citeauthoryear{{Katz}}{{Katz}}{2019}]{Katz2019}
{Katz} H.,  2019, in {Latif} M.,  {Schleicher} D.,  eds, , Formation of the
  First Black Holes.
pp 125--143, \mn@doi{10.1142/9789813227958\_0007}

\bibitem[\protect\citeauthoryear{{Katz}, {Sijacki}  \& {Haehnelt}}{{Katz}
  et~al.}{2015}]{Katz2015}
{Katz} H.,  {Sijacki} D.,   {Haehnelt} M.~G.,  2015, \mn@doi [\mnras]
  {10.1093/mnras/stv1048}, \href
  {https://ui.adsabs.harvard.edu/abs/2015MNRAS.451.2352K} {451, 2352}

\bibitem[\protect\citeauthoryear{{Klessen} \& {Hennebelle}}{{Klessen} \&
  {Hennebelle}}{2010}]{Klessen2010}
{Klessen} R.~S.,  {Hennebelle} P.,  2010, \mn@doi [\aap]
  {10.1051/0004-6361/200913780}, \href
  {https://ui.adsabs.harvard.edu/abs/2010A&A...520A..17K} {520, A17}

\bibitem[\protect\citeauthoryear{{Koushiappas}, {Bullock}  \&
  {Dekel}}{{Koushiappas} et~al.}{2004}]{Koushiappas2004}
{Koushiappas} S.~M.,  {Bullock} J.~S.,   {Dekel} A.,  2004, \mn@doi [\mnras]
  {10.1111/j.1365-2966.2004.08190.x}, \href
  {https://ui.adsabs.harvard.edu/abs/2004MNRAS.354..292K} {354, 292}

\bibitem[\protect\citeauthoryear{{Kroupa}, {Subr}, {Jerabkova}  \&
  {Wang}}{{Kroupa} et~al.}{2020a}]{Kroupa2020}
{Kroupa} P.,  {Subr} L.,  {Jerabkova} T.,   {Wang} L.,  2020a, \mn@doi [\mnras]
  {10.1093/mnras/staa2276}, \href
  {https://ui.adsabs.harvard.edu/abs/2020MNRAS.498.5652K} {498, 5652}

\bibitem[\protect\citeauthoryear{{Kroupa}, {Subr}, {Jerabkova}  \&
  {Wang}}{{Kroupa} et~al.}{2020b}]{Latif2014a}
{Kroupa} P.,  {Subr} L.,  {Jerabkova} T.,   {Wang} L.,  2020b, \mn@doi [\mnras]
  {10.1093/mnras/staa2276}, \href
  {https://ui.adsabs.harvard.edu/abs/2020MNRAS.498.5652K} {498, 5652}

\bibitem[\protect\citeauthoryear{{Latif} \& {Schleicher}}{{Latif} \&
  {Schleicher}}{2015a}]{LatifSchleicher2015}
{Latif} M.~A.,  {Schleicher} D.~R.~G.,  2015a, \mn@doi [\aap]
  {10.1051/0004-6361/201525855}, \href
  {https://ui.adsabs.harvard.edu/abs/2015A&A...578A.118L} {578, A118}

\bibitem[\protect\citeauthoryear{{Latif} \& {Schleicher}}{{Latif} \&
  {Schleicher}}{2015b}]{Latif2015}
{Latif} M.~A.,  {Schleicher} D.~R.~G.,  2015b, \mn@doi [\aap]
  {10.1051/0004-6361/201525855}, \href
  {https://ui.adsabs.harvard.edu/abs/2015A&A...578A.118L} {578, A118}

\bibitem[\protect\citeauthoryear{{Latif}, {Schleicher}, {Schmidt}  \&
  {Niemeyer}}{{Latif} et~al.}{2013a}]{Latif2013mag}
{Latif} M.~A.,  {Schleicher} D.~R.~G.,  {Schmidt} W.,   {Niemeyer} J.,  2013a,
  \mn@doi [\mnras] {10.1093/mnras/stt503}, \href
  {https://ui.adsabs.harvard.edu/abs/2013MNRAS.432..668L} {432, 668}

\bibitem[\protect\citeauthoryear{{Latif}, {Schleicher}, {Schmidt}  \&
  {Niemeyer}}{{Latif} et~al.}{2013b}]{Latif2013BH}
{Latif} M.~A.,  {Schleicher} D.~R.~G.,  {Schmidt} W.,   {Niemeyer} J.,  2013b,
  \mn@doi [\mnras] {10.1093/mnras/stt834}, \href
  {https://ui.adsabs.harvard.edu/abs/2013MNRAS.433.1607L} {433, 1607}

\bibitem[\protect\citeauthoryear{{Latif}, {Schleicher}  \& {Schmidt}}{{Latif}
  et~al.}{2014}]{Latif2014mag}
{Latif} M.~A.,  {Schleicher} D.~R.~G.,   {Schmidt} W.,  2014, \mn@doi [\mnras]
  {10.1093/mnras/stu357}, \href
  {https://ui.adsabs.harvard.edu/abs/2014MNRAS.440.1551L} {440, 1551}

\bibitem[\protect\citeauthoryear{{Latif}, {Omukai}, {Habouzit}, {Schleicher}
  \& {Volonteri}}{{Latif} et~al.}{2016}]{Latif2016}
{Latif} M.~A.,  {Omukai} K.,  {Habouzit} M.,  {Schleicher} D.~R.~G.,
  {Volonteri} M.,  2016, \mn@doi [\apj] {10.3847/0004-637X/823/1/40}, \href
  {https://ui.adsabs.harvard.edu/abs/2016ApJ...823...40L} {823, 40}

\bibitem[\protect\citeauthoryear{{Latif}, {Khochfar}  \& {Whalen}}{{Latif}
  et~al.}{2020}]{Latif2020}
{Latif} M.~A.,  {Khochfar} S.,   {Whalen} D.,  2020, \mn@doi [\apjl]
  {10.3847/2041-8213/ab7c61}, \href
  {https://ui.adsabs.harvard.edu/abs/2020ApJ...892L...4L} {892, L4}

\bibitem[\protect\citeauthoryear{{Latif}, {Khochfar}, {Schleicher}  \&
  {Whalen}}{{Latif} et~al.}{2021}]{Latif2021}
{Latif} M.~A.,  {Khochfar} S.,  {Schleicher} D.,   {Whalen} D.~J.,  2021,
  \mn@doi [\mnras] {10.1093/mnras/stab2708}, \href
  {https://ui.adsabs.harvard.edu/abs/2021MNRAS.508.1756L} {508, 1756}

\bibitem[\protect\citeauthoryear{{Latif}, {Whalen}  \& {Khochfar}}{{Latif}
  et~al.}{2022}]{Latif22}
{Latif} M.~A.,  {Whalen} D.,   {Khochfar} S.,  2022, \mn@doi [ApJ]
  {10.3847/1538-4357/ac3916}, \href
  {https://ui.adsabs.harvard.edu/abs/2022ApJ...925...28L} {925, 28}

\bibitem[\protect\citeauthoryear{{Lupi}, {Colpi}, {Devecchi}, {Galanti}  \&
  {Volonteri}}{{Lupi} et~al.}{2014}]{Lupi2014}
{Lupi} A.,  {Colpi} M.,  {Devecchi} B.,  {Galanti} G.,   {Volonteri} M.,  2014,
  \mn@doi [\mnras] {10.1093/mnras/stu1120}, \href
  {https://ui.adsabs.harvard.edu/abs/2014MNRAS.442.3616L} {442, 3616}

\bibitem[\protect\citeauthoryear{{Mac Low}}{{Mac Low}}{1999}]{MacLow1999}
{Mac Low} M.-M.,  1999, \mn@doi [\apj] {10.1086/307784}, \href
  {https://ui.adsabs.harvard.edu/abs/1999ApJ...524..169M} {524, 169}

\bibitem[\protect\citeauthoryear{{Maccarone} \& {Zurek}}{{Maccarone} \&
  {Zurek}}{2012}]{Maccarone2012}
{Maccarone} T.~J.,  {Zurek} D.~R.,  2012, \mn@doi [\mnras]
  {10.1111/j.1365-2966.2011.20328.x}, \href
  {https://ui.adsabs.harvard.edu/abs/2012MNRAS.423....2M} {423, 2}

\bibitem[\protect\citeauthoryear{{Matsukoba}, {Vorobyov}, {Sugimura}, {Chon},
  {Hosokawa}  \& {Omukai}}{{Matsukoba} et~al.}{2021}]{Chon2021}
{Matsukoba} R.,  {Vorobyov} E.~I.,  {Sugimura} K.,  {Chon} S.,  {Hosokawa} T.,
   {Omukai} K.,  2021, \mn@doi [\mnras] {10.1093/mnras/staa3462}, \href
  {https://ui.adsabs.harvard.edu/abs/2021MNRAS.500.4126M} {500, 4126}

\bibitem[\protect\citeauthoryear{{Nagele}, {Umeda}, {Takahashi}, {Yoshida}  \&
  {Sumiyoshi}}{{Nagele} et~al.}{2020}]{Nagele2020}
{Nagele} C.,  {Umeda} H.,  {Takahashi} K.,  {Yoshida} T.,   {Sumiyoshi} K.,
  2020, \mn@doi [\mnras] {10.1093/mnras/staa1636}, \href
  {https://ui.adsabs.harvard.edu/abs/2020MNRAS.496.1224N} {496, 1224}

\bibitem[\protect\citeauthoryear{{Omukai}, {Schneider}  \& {Haiman}}{{Omukai}
  et~al.}{2008}]{Omukai2008}
{Omukai} K.,  {Schneider} R.,   {Haiman} Z.,  2008, \mn@doi [\apj]
  {10.1086/591636}, \href
  {https://ui.adsabs.harvard.edu/abs/2008ApJ...686..801O} {686, 801}

\bibitem[\protect\citeauthoryear{{Ostriker}}{{Ostriker}}{1999}]{Ostriker1999}
{Ostriker} E.~C.,  1999, \mn@doi [\apj] {10.1086/306858}, \href
  {https://ui.adsabs.harvard.edu/abs/1999ApJ...513..252O} {513, 252}

\bibitem[\protect\citeauthoryear{{Peters}, {Schleicher}, {Smith}, {Schmidt}  \&
  {Klessen}}{{Peters} et~al.}{2014}]{Peters2014}
{Peters} T.,  {Schleicher} D. R.~G.,  {Smith} R.~J.,  {Schmidt} W.,   {Klessen}
  R.~S.,  2014, \mn@doi [\mnras] {10.1093/mnras/stu1097}, \href
  {https://ui.adsabs.harvard.edu/abs/2014MNRAS.442.3112P} {442, 3112}

\bibitem[\protect\citeauthoryear{{Pfister}, {Volonteri}, {Dubois}, {Dotti}  \&
  {Colpi}}{{Pfister} et~al.}{2019}]{Pfister2019}
{Pfister} H.,  {Volonteri} M.,  {Dubois} Y.,  {Dotti} M.,   {Colpi} M.,  2019,
  \mn@doi [\mnras] {10.1093/mnras/stz822}, \href
  {https://ui.adsabs.harvard.edu/abs/2019MNRAS.486..101P} {486, 101}

\bibitem[\protect\citeauthoryear{{Portegies Zwart} \& {McMillan}}{{Portegies
  Zwart} \& {McMillan}}{2002}]{Poregies2002}
{Portegies Zwart} S.~F.,  {McMillan} S. L.~W.,  2002, \mn@doi [\apj]
  {10.1086/341798}, \href
  {https://ui.adsabs.harvard.edu/abs/2002ApJ...576..899P} {576, 899}

\bibitem[\protect\citeauthoryear{{Rahner}, {Pellegrini}, {Glover}  \&
  {Klessen}}{{Rahner} et~al.}{2017}]{Rahner2017}
{Rahner} D.,  {Pellegrini} E.~W.,  {Glover} S. C.~O.,   {Klessen} R.~S.,  2017,
  \mn@doi [\mnras] {10.1093/mnras/stx1532}, \href
  {https://ui.adsabs.harvard.edu/abs/2017MNRAS.470.4453R} {470, 4453}

\bibitem[\protect\citeauthoryear{{Rahner}, {Pellegrini}, {Glover}  \&
  {Klessen}}{{Rahner} et~al.}{2019}]{Rahner2019}
{Rahner} D.,  {Pellegrini} E.~W.,  {Glover} S. C.~O.,   {Klessen} R.~S.,  2019,
  \mn@doi [\mnras] {10.1093/mnras/sty3295}, \href
  {https://ui.adsabs.harvard.edu/abs/2019MNRAS.483.2547R} {483, 2547}

\bibitem[\protect\citeauthoryear{{Rees}}{{Rees}}{1984}]{Rees1984}
{Rees} M.~J.,  1984, \mn@doi [\araa] {10.1146/annurev.aa.22.090184.002351},
  \href {https://ui.adsabs.harvard.edu/abs/1984ARA&A..22..471R} {22, 471}

\bibitem[\protect\citeauthoryear{{Regan}, {Visbal}, {Wise}, {Haiman},
  {Johansson}  \& {Bryan}}{{Regan} et~al.}{2017}]{Regan2017}
{Regan} J.~A.,  {Visbal} E.,  {Wise} J.~H.,  {Haiman} Z.,  {Johansson} P.~H.,
  {Bryan} G.~L.,  2017, \mn@doi [Nature Astronomy] {10.1038/s41550-017-0075},
  \href {https://ui.adsabs.harvard.edu/abs/2017NatAs...1E..75R} {1, 0075}

\bibitem[\protect\citeauthoryear{{Regan}, {Wise}, {Woods}, {Downes}, {O'Shea}
  \& {Norman}}{{Regan} et~al.}{2020}]{Regan2020}
{Regan} J.~A.,  {Wise} J.~H.,  {Woods} T.~E.,  {Downes} T.~P.,  {O'Shea} B.~W.,
    {Norman} M.~L.,  2020, \mn@doi [The Open Journal of Astrophysics]
  {10.21105/astro.2008.08090}, \href
  {https://ui.adsabs.harvard.edu/abs/2020OJAp....3E..15R} {3, 15}

\bibitem[\protect\citeauthoryear{{Reinoso}, {Schleicher}, {Fellhauer},
  {Klessen}  \& {Boekholt}}{{Reinoso} et~al.}{2018}]{Reinoso2018}
{Reinoso} B.,  {Schleicher} D.~R.~G.,  {Fellhauer} M.,  {Klessen} R.~S.,
  {Boekholt} T.~C.~N.,  2018, \mn@doi [\aap] {10.1051/0004-6361/201732224},
  \href {https://ui.adsabs.harvard.edu/abs/2018A&A...614A..14R} {614, A14}

\bibitem[\protect\citeauthoryear{{Reinoso}, {Schleicher}, {Fellhauer}, {Leigh}
  \& {Klessen}}{{Reinoso} et~al.}{2020}]{Reinoso2020}
{Reinoso} B.,  {Schleicher} D.~R.~G.,  {Fellhauer} M.,  {Leigh} N.~W.~C.,
  {Klessen} R.~S.,  2020, \mn@doi [\aap] {10.1051/0004-6361/202037843}, \href
  {https://ui.adsabs.harvard.edu/abs/2020A&A...639A..92R} {639, A92}

\bibitem[\protect\citeauthoryear{{Sakurai}, {Yoshida}, {Fujii}  \&
  {Hirano}}{{Sakurai} et~al.}{2017}]{Sakurai2017}
{Sakurai} Y.,  {Yoshida} N.,  {Fujii} M.~S.,   {Hirano} S.,  2017, \mn@doi
  [\mnras] {10.1093/mnras/stx2044}, \href
  {https://ui.adsabs.harvard.edu/abs/2017MNRAS.472.1677S} {472, 1677}

\bibitem[\protect\citeauthoryear{{Sakurai}, {Yoshida}  \& {Fujii}}{{Sakurai}
  et~al.}{2019}]{Sakurai2019}
{Sakurai} Y.,  {Yoshida} N.,   {Fujii} M.~S.,  2019, \mn@doi [\mnras]
  {10.1093/mnras/stz315}, \href
  {https://ui.adsabs.harvard.edu/abs/2019MNRAS.484.4665S} {484, 4665}

\bibitem[\protect\citeauthoryear{{Sassano}, {Schneider}, {Valiante},
  {Inayoshi}, {Chon}, {Omukai}, {Mayer}  \& {Capelo}}{{Sassano}
  et~al.}{2021}]{Sassano2021}
{Sassano} F.,  {Schneider} R.,  {Valiante} R.,  {Inayoshi} K.,  {Chon} S.,
  {Omukai} K.,  {Mayer} L.,   {Capelo} P.~R.,  2021, \mn@doi [\mnras]
  {10.1093/mnras/stab1737}, \href
  {https://ui.adsabs.harvard.edu/abs/2021MNRAS.506..613S} {506, 613}

\bibitem[\protect\citeauthoryear{{Schaerer}}{{Schaerer}}{2002}]{Schaerer2002}
{Schaerer} D.,  2002, \mn@doi [\aap] {10.1051/0004-6361:20011619}, \href
  {https://ui.adsabs.harvard.edu/abs/2002A&A...382...28S} {382, 28}

\bibitem[\protect\citeauthoryear{{Schleicher}, {Spaans}  \&
  {Glover}}{{Schleicher} et~al.}{2010}]{Schleicher10b}
{Schleicher} D.~R.~G.,  {Spaans} M.,   {Glover} S.~C.~O.,  2010, \mn@doi [ApJL]
  {10.1088/2041-8205/712/1/L69}, \href
  {http://adsabs.harvard.edu/abs/2010ApJ...712L..69S} {712, L69}

\bibitem[\protect\citeauthoryear{{Schleicher}, {Schober}, {Federrath}, {Bovino}
   \& {Schmidt}}{{Schleicher} et~al.}{2013a}]{Schleicher2013mag}
{Schleicher} D. R.~G.,  {Schober} J.,  {Federrath} C.,  {Bovino} S.,
  {Schmidt} W.,  2013a, \mn@doi [New Journal of Physics]
  {10.1088/1367-2630/15/2/023017}, \href
  {https://ui.adsabs.harvard.edu/abs/2013NJPh...15b3017S} {15, 023017}

\bibitem[\protect\citeauthoryear{{Schleicher}, {Palla}, {Ferrara}, {Galli}  \&
  {Latif}}{{Schleicher} et~al.}{2013b}]{Schleicher2013}
{Schleicher} D. R.~G.,  {Palla} F.,  {Ferrara} A.,  {Galli} D.,   {Latif} M.,
  2013b, \mn@doi [\aap] {10.1051/0004-6361/201321949}, \href
  {https://ui.adsabs.harvard.edu/abs/2013A&A...558A..59S} {558, A59}

\bibitem[\protect\citeauthoryear{{Schneider} \& {Omukai}}{{Schneider} \&
  {Omukai}}{2010}]{Schneider2010}
{Schneider} R.,  {Omukai} K.,  2010, \mn@doi [\mnras]
  {10.1111/j.1365-2966.2009.15891.x}, \href
  {https://ui.adsabs.harvard.edu/abs/2010MNRAS.402..429S} {402, 429}

\bibitem[\protect\citeauthoryear{{Schneider}, {Omukai}, {Inoue}  \&
  {Ferrara}}{{Schneider} et~al.}{2006}]{Schneider2006}
{Schneider} R.,  {Omukai} K.,  {Inoue} A.~K.,   {Ferrara} A.,  2006, \mn@doi
  [\mnras] {10.1111/j.1365-2966.2006.10391.x}, \href
  {https://ui.adsabs.harvard.edu/abs/2006MNRAS.369.1437S} {369, 1437}

\bibitem[\protect\citeauthoryear{{Schober}, {Schleicher}, {Federrath}, {Bovino}
   \& {Klessen}}{{Schober} et~al.}{2015}]{Schober2015}
{Schober} J.,  {Schleicher} D.~R.~G.,  {Federrath} C.,  {Bovino} S.,
  {Klessen} R.~S.,  2015, \mn@doi [\pre] {10.1103/PhysRevE.92.023010}, \href
  {https://ui.adsabs.harvard.edu/abs/2015PhRvE..92b3010S} {92, 023010}

\bibitem[\protect\citeauthoryear{{Shapiro}}{{Shapiro}}{2005}]{Shapiro2005}
{Shapiro} S.~L.,  2005, \mn@doi [\apj] {10.1086/427065}, \href
  {https://ui.adsabs.harvard.edu/abs/2005ApJ...620...59S} {620, 59}

\bibitem[\protect\citeauthoryear{{Sharda}, {Federrath}, {Krumholz}  \&
  {Schleicher}}{{Sharda} et~al.}{2021}]{Sharda2021}
{Sharda} P.,  {Federrath} C.,  {Krumholz} M.~R.,   {Schleicher} D. R.~G.,
  2021, \mn@doi [\mnras] {10.1093/mnras/stab531}, \href
  {https://ui.adsabs.harvard.edu/abs/2021MNRAS.503.2014S} {503, 2014}

\bibitem[\protect\citeauthoryear{{Spitzer}}{{Spitzer}}{1987}]{Spitzer1987}
{Spitzer} L.,  1987, {Dynamical evolution of globular clusters}

\bibitem[\protect\citeauthoryear{{Str{\"o}mgren}}{{Str{\"o}mgren}}{1939}]{Stromgren1939}
{Str{\"o}mgren} B.,  1939, \mn@doi [\apj] {10.1086/144074}, \href
  {https://ui.adsabs.harvard.edu/abs/1939ApJ....89..526S} {89, 526}

\bibitem[\protect\citeauthoryear{{Suazo}, {Prieto}, {Escala}  \&
  {Schleicher}}{{Suazo} et~al.}{2019}]{Suazo2019}
{Suazo} M.,  {Prieto} J.,  {Escala} A.,   {Schleicher} D. R.~G.,  2019, \mn@doi
  [\apj] {10.3847/1538-4357/ab45eb}, \href
  {https://ui.adsabs.harvard.edu/abs/2019ApJ...885..127S} {885, 127}

\bibitem[\protect\citeauthoryear{{Sugimura}, {Omukai}  \& {Inoue}}{{Sugimura}
  et~al.}{2014}]{Sugimura2014}
{Sugimura} K.,  {Omukai} K.,   {Inoue} A.~K.,  2014, \mn@doi [\mnras]
  {10.1093/mnras/stu1778}, \href
  {https://ui.adsabs.harvard.edu/abs/2014MNRAS.445..544S} {445, 544}

\bibitem[\protect\citeauthoryear{{Sun}, {Ruiz}  \& {Shapiro}}{{Sun}
  et~al.}{2018}]{Sun2018}
{Sun} L.,  {Ruiz} M.,   {Shapiro} S.~L.,  2018, \mn@doi [\prd]
  {10.1103/PhysRevD.98.103008}, \href
  {https://ui.adsabs.harvard.edu/abs/2018PhRvD..98j3008S} {98, 103008}

\bibitem[\protect\citeauthoryear{{Tagawa}, {Haiman}  \& {Kocsis}}{{Tagawa}
  et~al.}{2020}]{Tagawa2020}
{Tagawa} H.,  {Haiman} Z.,   {Kocsis} B.,  2020, \mn@doi [\apj]
  {10.3847/1538-4357/ab7922}, \href
  {https://ui.adsabs.harvard.edu/abs/2020ApJ...892...36T} {892, 36}

\bibitem[\protect\citeauthoryear{{Trinca}, {Schneider}, {Valiante}, {Graziani},
  {Zappacosta}  \& {Shankar}}{{Trinca} et~al.}{2022}]{Trinca2022}
{Trinca} A.,  {Schneider} R.,  {Valiante} R.,  {Graziani} L.,  {Zappacosta} L.,
    {Shankar} F.,  2022, \mn@doi [\mnras] {10.1093/mnras/stac062}, \href
  {https://ui.adsabs.harvard.edu/abs/2022MNRAS.511..616T} {511, 616}

\bibitem[\protect\citeauthoryear{{Uchida}, {Shibata}, {Yoshida}, {Sekiguchi}
  \& {Umeda}}{{Uchida} et~al.}{2017}]{Uchida2017}
{Uchida} H.,  {Shibata} M.,  {Yoshida} T.,  {Sekiguchi} Y.,   {Umeda} H.,
  2017, \mn@doi [\prd] {10.1103/PhysRevD.96.083016}, \href
  {https://ui.adsabs.harvard.edu/abs/2017PhRvD..96h3016U} {96, 083016}

\bibitem[\protect\citeauthoryear{{Umeda}, {Hosokawa}, {Omukai}  \&
  {Yoshida}}{{Umeda} et~al.}{2016}]{Umeda2016}
{Umeda} H.,  {Hosokawa} T.,  {Omukai} K.,   {Yoshida} N.,  2016, \mn@doi
  [\apjl] {10.3847/2041-8205/830/2/L34}, \href
  {https://ui.adsabs.harvard.edu/abs/2016ApJ...830L..34U} {830, L34}

\bibitem[\protect\citeauthoryear{{Valiante}, {Schneider}, {Volonteri}  \&
  {Omukai}}{{Valiante} et~al.}{2016}]{Valiante2016}
{Valiante} R.,  {Schneider} R.,  {Volonteri} M.,   {Omukai} K.,  2016, \mn@doi
  [\mnras] {10.1093/mnras/stw225}, \href
  {https://ui.adsabs.harvard.edu/abs/2016MNRAS.457.3356V} {457, 3356}

\bibitem[\protect\citeauthoryear{{Vergara}, {Schleicher}, {Boekholt},
  {Reinoso}, {Fellhauer}, {Klessen}  \& {Leigh}}{{Vergara}
  et~al.}{2021}]{Vergara2021}
{Vergara} M.~Z.~C.,  {Schleicher} D.~R.~G.,  {Boekholt} T.~C.~N.,  {Reinoso}
  B.,  {Fellhauer} M.,  {Klessen} R.~S.,   {Leigh} N.~W.~C.,  2021, \mn@doi
  [\aap] {10.1051/0004-6361/202140298}, \href
  {https://ui.adsabs.harvard.edu/abs/2021A&A...649A.160V} {649, A160}

\bibitem[\protect\citeauthoryear{{Wang} et~al.,}{{Wang}
  et~al.}{2021}]{Wang2021}
{Wang} F.,  et~al., 2021, \mn@doi [\apj] {10.3847/1538-4357/abcc5e}, \href
  {https://ui.adsabs.harvard.edu/abs/2021ApJ...908...53W} {908, 53}

\bibitem[\protect\citeauthoryear{{Windhorst} et~al.,}{{Windhorst}
  et~al.}{2018}]{Windhorst2018}
{Windhorst} R.~A.,  et~al., 2018, \mn@doi [\apjs] {10.3847/1538-4365/aaa760},
  \href {https://ui.adsabs.harvard.edu/abs/2018ApJS..234...41W} {234, 41}

\bibitem[\protect\citeauthoryear{{Wise}, {Turk}  \& {Abel}}{{Wise}
  et~al.}{2008}]{Wise2008a}
{Wise} J.~H.,  {Turk} M.~J.,   {Abel} T.,  2008, \mn@doi [ApJ]
  {10.1086/588209}, \href {http://adsabs.harvard.edu/abs/2008ApJ...682..745W}
  {682, 745}

\bibitem[\protect\citeauthoryear{{Woods}, {Heger}, {Whalen}, {Haemmerl{\'e}}
  \& {Klessen}}{{Woods} et~al.}{2017}]{Woods2017}
{Woods} T.~E.,  {Heger} A.,  {Whalen} D.~J.,  {Haemmerl{\'e}} L.,   {Klessen}
  R.~S.,  2017, \mn@doi [\apjl] {10.3847/2041-8213/aa7412}, \href
  {https://ui.adsabs.harvard.edu/abs/2017ApJ...842L...6W} {842, L6}

\makeatother
\end{thebibliography}





\bsp	
\label{lastpage}
\end{document}